\documentclass[%
 reprint,
 amsmath,amssymb,
 aps,prd,nofootinbib,showpacs
]{revtex4-1}

\usepackage{graphicx}
\usepackage{dcolumn}
\usepackage{bm}
\usepackage{color}
\usepackage{amsmath}
\usepackage{epstopdf}
\usepackage{tabularx,ragged2e,booktabs}

\usepackage{amssymb}
\usepackage{fullpage}
\usepackage{hyperref}

\newcolumntype{C}[1]{>{\Centering}m{#1}}

\newcommand{\lsup}[2]{ \ensuremath{{}^{#2}\!{#1}}}

\newcommand{\Beq}{\begin{equation}\begin{aligned}}
\newcommand{\Eeq}{\end{aligned}\end{equation}}

\newcommand{\lsim}{\mathrel{\hbox{\rlap{\lower.55ex\hbox{$\sim$}} \kern-.3em \raise.4ex \hbox{$<$}}}}
\newcommand{\gsim}{\mathrel{\hbox{\rlap{\lower.55ex\hbox{$\sim$}} \kern-.3em \raise.4ex \hbox{$>$}}}}

\begin{document}

\title{Probing a panoply of curvaton-decay scenarios using CMB data}

\author{Tristan L.~Smith$^1$}\email{tsmith2@swarthmore.edu}
\author{Daniel Grin$^{2}$}\email{dgrin@kicp.uchicago.edu}
\affiliation{$^1$Swarthmore College, Department of Physics and Astronomy, Swarthmore, PA 19081, USA}
\affiliation{$^{2}$Department of Astronomy \& Astrophysics, Kavli Institute for Cosmological Physics, University of Chicago, Chicago, IL 60637, U.S.A}

\date{\today}
\begin{abstract}
In the curvaton scenario, primordial curvature perturbations are produced by a second field that is sub-dominant during inflation. Depending on how the curvaton decays [possibly producing baryon number, lepton number, or cold dark matter (CDM)], mixtures of correlated isocurvature perturbations are produced, allowing the curvaton scenario to be tested using cosmic microwave background (CMB) data. Here, a full range of $27$ curvaton-decay scenarios is compared with CMB data, placing limits on the curvaton fraction at decay, $r_{D}$, and the lepton asymmetry, $\xi_{\rm lep}$. If baryon number is generated by curvaton decay and CDM before (or vice-versa), these limits imply specific predictions for non-Gaussian signatures testable by future CMB experiments and upcoming large-scale-structure surveys.
 \end{abstract}
\pacs{95.35.+d, 98.80.Cq,98.70.Vc,98.80.-k}
               
\maketitle
\section{Introduction}\label{sec:intro}
The observed cosmic microwave background (CMB) anisotropies and large-scale structure of the universe are thought to result from primordial curvature perturbations. The prevailing model is that these perturbations are produced during inflation, an epoch of accelerated cosmological expansion preceding the radiation-dominated era. In the simplest scenarios, both the accelerated expansion and the curvature perturbations result from the dynamics of a single field (the \emph{inflaton}) \cite{Guth:1982ec,Bardeen:1983qw,Mukhanov:1990me}. At the end of inflation, the inflaton field is thought to decay and initiate the radiation-dominated era, a process known as reheating \cite{Kofman:1994rk,Amin:2014eta}.

Standard single-field models of inflation produce nearly scale-invariant, Gaussian, and adiabatic primordial fluctuations \cite{Bardeen:1983qw,Mukhanov:1990me}. It may be challenging for the dynamics of a single field to satisfy observational constraints to the amplitude and scale-dependence of the curvature perturbations as well as constraints to the amplitude of a background of primordial gravitational waves \cite{Lyth:2001nq,Lyth:2002my}. In order to ease these requirements, a second field (the \emph{curvaton}) could source curvature perturbations and later decay \cite{Mollerach:1989hu,Mukhanov:1990me,Moroi:2001ct,Lyth:2001nq,Lyth:2002my}. There are a variety of candidates for the curvaton motivated by high-energy particle theory \cite{Postma:2002et,Kasuya:2003va,Ikegami:2004ve,Mazumdar:2004qv,Allahverdi:2006dr,Papantonopoulos:2006xi,Mazumdar:2010sa,Mazumdar:2011xe}. In the curvaton scenario, constraints are more permissive because the inflaton need only produce a sufficiently long epoch of acceleration to dilute topological defects and does not have to be the main source of perturbations \cite{Mollerach:1989hu,Mukhanov:1990me,Lyth:2001nq,Lyth:2002my,Enqvist:2009zf}.
\begin{figure}[ht]
\includegraphics[width=3 in]{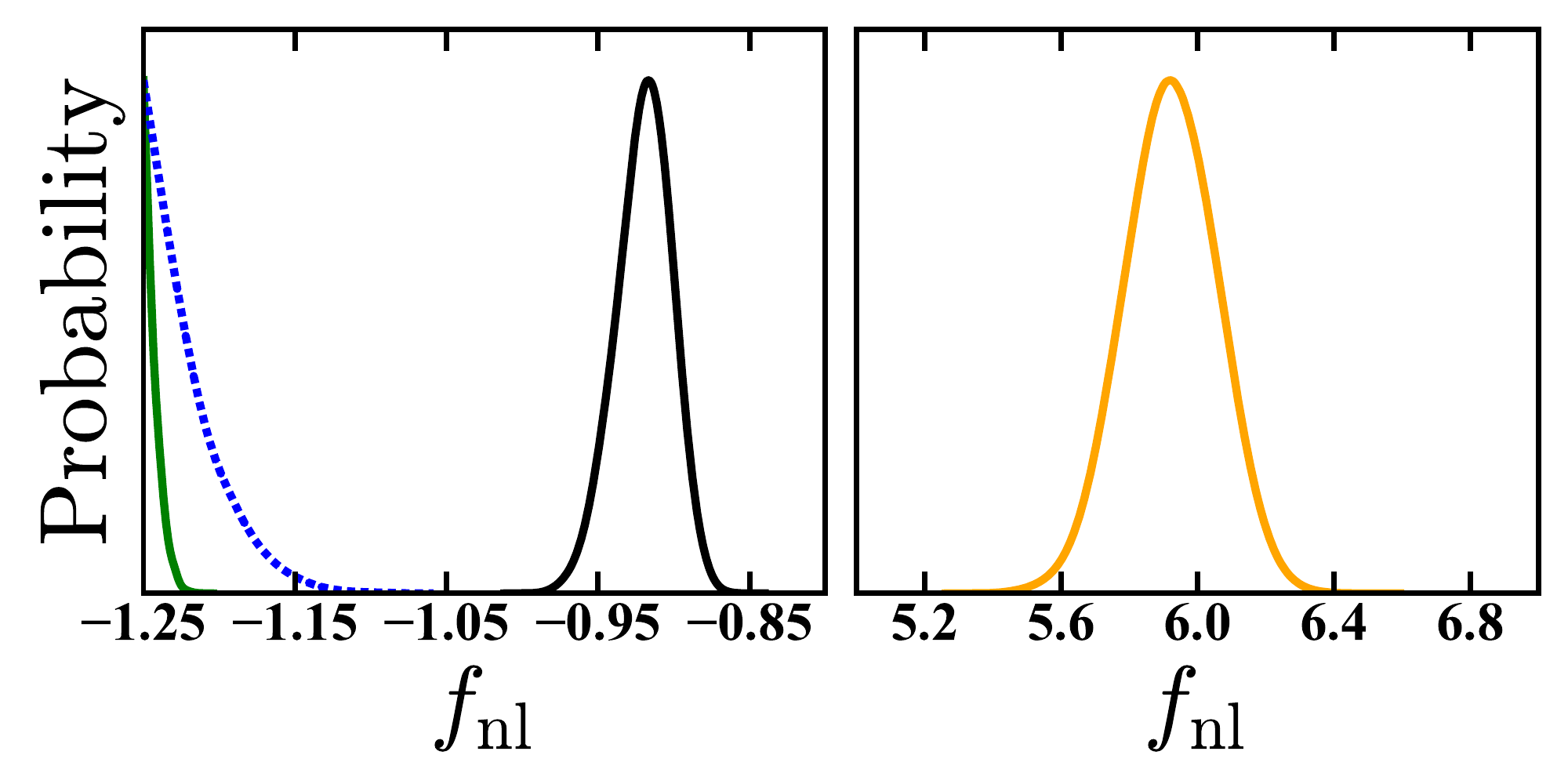}
\caption{Prediction for the amplitude $f_{\rm nl}$ of primordial non-Gaussianity in curvaton-decay scenarios allowed by isocurvature constraints. The left panel shows models with $f_{\rm nl}\sim 1$, which are potentially testable by future high-redshift 21-cm surveys \cite{Cooray:2004kt,Pillepich:2006fj,Munoz:2015eqa}. The solid green curve shows the case in which baryon number/CDM are generated after/by curvaton decay. The dotted blue curve shows the case in which baryon number/CDM are generated by/after curvaton decay. The solid black curve shows the case in which baryon number/CDM are generated before/by curvaton decay. Right panel shows the predicted $f_{\rm nl}$ values if baryon number/CDM are generated by/before curvaton decay, which could be tested using scale-dependent bias measurements from future galaxy surveys with sensitivity $\Delta f_{\rm nl}\simeq 1$. }
\label{fig:fnl_all}
\end{figure} 

This scenario is distinct from single-field models in predicting a non-adiabatic and non-Gaussian component to primordial fluctuations  \cite{Linde:1984ti, Linde:1996gt, Lyth:2002my,Moroi:2002rd,Lyth:2003ip,Malik:2006pm,Enqvist:2009zf,Kawasaki:2011pd,Vennin:2015vfa}. Depending on when the curvaton decays relative to the production of baryon number, lepton number, and cold dark matter (CDM), primordial fluctuations in different species could carry an isocurvature component, in which the relative number densities of different species fluctuate in space. In the simplest models, these isocurvature fluctuations are totally correlated (or anti-correlated) with the dominant adiabatic component. Curvaton density fluctuations are non-Gaussian, and so the curvature perturbation is non-Gaussian \cite{Linde:1996gt,Malik:2006pm}.

The level of non-Gaussianity is set by $r_{D}$, a parameter describing the curvaton energy-density. The level of isocurvature is set by $r_{D}$ and $\xi_{\rm lep}$, the chemical potential describing cosmological lepton number \cite{Lyth:2002my}. Both parameters are constrained by observations. 

Isocurvature perturbations alter the phase-structure and large-scale amplitude of CMB power spectra \cite{Bond:1984fp,Kodama:1986fg,Kodama:1986ud,Hu:1994jd,Moodley:2004nz,Bean:2006qz}. \textit{Planck} satellite observations thus indicate that CMB anisotropy power spectra are consistent with adiabatic fluctuations, requiring that isocurvature fluctuations contribute a fraction $\lsim 10^{-3}-0.1$ of the total observed power, depending on various assumptions \cite{DiValentino:2014eea,Ade:2015lrj,Ade:2015xua}. Big-bang nucleosynthesis abundances are altered if $\xi_{\rm lep}^{2}> 0$, and so the primordial $\lsup{{\rm He}}{4}$ and deuterium abundances impose the limit $|\xi_{\rm lep}|\leq 0.03$ \cite{Dolgov:2002ab,Cuoco:2003cu,Guzzo:2003xk,Mangano:2011ip,DiValentino:2014eea}.

In past work comparing curvaton model-predictions with CMB data, isocurvature constraints were obtained considering a single mode (neutrino, CDM, or baryon) at a time, with consideration limited to several curvaton decay-scenarios \cite{Gordon:2002gv,Gordon:2009wx,DiValentino:2011sv,DiValentino:2014eea,Ade:2015lrj,Ade:2015xua}. Priors and parameter-space exploration were implemented on the cross/auto-power spectrum amplitudes and correlation coefficients of single isocurvature modes, rather than $r_{D}$ and $\xi_{\rm lep}$, and then mapped to the curvaton parameter space. Neutrino isocurvature perturbations were not included.\footnote{Exceptions are Refs.~\cite{Cuoco:2003cu,DiValentino:2011sv,DiValentino:2014eea}, which included isocurvature in neutrinos but not other species.}

In fact, each curvaton-decay scenario makes specific predictions for the amplitudes and cross-correlations (with $\zeta$) of each isocurvature mode: the baryon isocurvature mode, the CDM isocurvature mode, and the neutrino isocurvature density mode \cite{Lyth:2001nq,Lyth:2003ip,Gordon:2003hw}. We take a different approach and separately consider all $27$ curvaton-decay scenarios. We use $2015$ \textit{Planck} CMB temperature and polarization data to determine the allowed parameter space of $r_{D}$ and $\xi_{\rm lep}$ (breaking degeneracies with other data), computing the full set of isocurvature mode amplitudes and cross-correlation spectra (with $\zeta$)  for each set of parameter values. We use a Monte Carlo Markov Chain (MCMC) analysis to obtain constraints to all these scenarios. We also perform a Fisher-matrix analysis to determine the sensitivity of a future cosmic-variance limited experiment to these curvaton-decay scenarios.

The models fall into several categories. Some decay scenarios generate purely adiabatic perturbations, and these are always allowed, and these are unconstrained by limits to isocurvature perturbations. Some generate order unity isocurvature fluctuations between non-relativistic matter and radiation, independent of $r_{D}$ and $\xi_{\rm lep}$ values, and these are not allowed by the CMB data. Others generate isocurvature perturbations that vanish when $r_{D}=1$. Here, the data impose lower limits to $r_{D}$, with $95\%$-confidence regions given by $r_{D}>0.93-0.99$, depending on precise model assumptions. 

Finally, two cases lead to non-zero isocurvature perturbations in both the baryon and CDM. The only way for these scenarios to agree with the CMB data is for the baryon and CDM isocurvature modes have opposite sign and nearly equal amplitudes, producing what is known as a compensated isocurvature perturbation \cite{Grin:2011tf}.  This naturally leads to a measured value of $r_D$ which is significantly different from unity. For the curvaton-decay scenario in which baryon number/CDM are generated by/before curvaton decay, we find that $r_{\rm D}=0.1602^{+.0051}_{-0.0047}$, while for the scenario in which baryon number/CDM are generated before/by curvaton decay, $r_{\rm D}=0.8492^{+0.0099}_{-0.0096}$.

All of these decay scenarios (except the one where both CDM and baryon number are produced after curvaton decay) make specific predictions for the amplitude $f_{\rm nl}$ of local-type primordial non-Gaussianity, shown by the distributions in Fig.~\ref{fig:fnl_all}. These are all consistent with \textit{Planck} limits to $f_{\rm nl}$ \cite{Ade:2015ava}. Future measurements of scale-dependent bias in galaxy surveys (with sensitivity $\Delta f_{\rm nl}\simeq \pm 1$) \cite{Dalal:2007cu,dePutter:2014lna} and high-redshift $21$-cm surveys (with sensitivity $\Delta f_{\rm nl}\simeq \pm 0.03$) \cite{Cooray:2004kt,Pillepich:2006fj,Munoz:2015eqa} could rule out these decay scenarios.

We begin in Sec.~\ref{sec:curv_model} by reviewing basic aspects of the curvaton model, including the production of curvature and isocurvature perturbations. In Sec.~\ref{sec:decay_scen} we continue with a detailed discussion of curvaton-decay scenarios and the resulting mixtures of curvature and isocurvature fluctuations.  The data sets, methodology, and resulting constraints on these scenarios are presented in Sec.~\ref{sec:data}. We present our conclusions in Sec.~\ref{sec:conclusions}.

\section{The curvaton model}
\label{sec:curv_model}

The family of inflationary models is extremely rich. Nonetheless, a successful inflationary model must meet some fairly stringent requirements, producing a sufficient number ($\sim 60$) of e-foldings to dilute dangerous early relics, generating the observed value of $A_{s}=2.2\times 10^{-9}$, and agreeing with ever more precise measurements of the scalar spectral index $n_{s}\simeq 0.96$ \cite{Ade:2015lrj}. Limits to the tensor-to-scalar ratio ($r<0.11$ \cite{Ade:2015lrj}) must also be met. If these limits turn into detections, single-field slow-roll models must further obey a consistency relation, $r=16\epsilon$ (see Ref.~\cite{Stewart:1993bc} and references therein), which relates  $r$ to the slow-roll parameter $\epsilon$. In fact, current data already rules out the simplest of inflationary models \cite{Ade:2015lrj}.

One alternative to simple inflationary models is the \textit{curvaton} scenario, in which the inflaton ($\phi$) drives exponential cosmic expansion but is not the primary source of the observed cosmological fluctuations. Instead, a sub-dominant spectator field, the curvaton ($\sigma$), acquires quantum fluctuations that are frozen after $\sigma$ perturbation modes cross the horizon during inflation. The curvaton field then has a dimensionless fluctuation power-spectrum of \cite{Lyth:2002my,Gordon:2002gv,Gupta:2003jc,Enqvist:2009zf}
\begin{equation}
\Delta^{2}_{\sigma\sigma}(k)=\left(\frac{H_{I}}{2\pi}\right)_{k=aH}^{2},\end{equation} where $H_{I}$ is the inflationary Hubble parameter when the mode with wave-number $k$ freezes out. Initially, these fluctuations are isocurvature perturbations, as the curvaton is energetically sub-dominant to the thermal bath (with energy-density $\rho_{R}$) produced at the end of inflation \cite{Linde:1996gt,Lyth:2002my,Mollerach:1989hu,Mukhanov:1990me,Moroi:2001ct,Lyth:2003ip}. The curvaton has mass $m_{\sigma}$, and once the condition $m_{\sigma}\gg 3H$ is met (where $H$ is the Hubble parameter), $\sigma$ begins to coherently oscillate. The curvaton energy-density then redshifts as $\rho_{\sigma}\simeq a^{-3}$, where $a$ its the cosmological scale factor. As the scaling $\rho_{\sigma}\sim a^{-3}$ is slower than $\rho_{R}\sim a^{-4}$, the curvaton becomes increasingly energetically important, converting the initial isocurvature fluctuation into a gauge-invariant curvature perturbation $\zeta$ \cite{Lyth:2002my,Gordon:2002gv,Gupta:2003jc,Enqvist:2009zf}.  Eventually the curvaton decays, initating the usual epoch of radiation domination. 

During radiation domination, the cosmic equation of state is constant, and it can be shown that this implies conservation of super-horizon modes of $\zeta$, with value\begin{equation}
\zeta=\left(1-r_{D}\right)\left.\zeta_{\phi}\right|_{D}+r_{D}\left.\zeta_{\sigma}\right|_{D}\label{eq:zeta_total},
\end{equation}where 
\begin{equation}
r_{D}=\left.\frac{\rho_{\sigma}}{\left(\rho_{\sigma}+4\rho_{R}/3\right)}\right|_{D},\end{equation} is the fractional contribution of the curvaton to the trace of the stress-energy tensor just before curvaton decay. Here $\zeta_{x}$ denotes the spatial-curvature perturbation on hypersurfaces of constant $x$ energy density (or equivalently, the energy density perturbation on surfaces of constant total $\zeta$). The notation $\left.\zeta_{x}\right|_{D}$ indicates that $\zeta_{x}$ is evaluated at the moment of curvaton decay.  For the duration of this paper, we neglect the  time-dependence of the curvaton-decay rate \cite{Kitajima:2014xna} and assume the usual `instantaneous-decay approximation' that the curvaton decays instantaneously. 

In principle, as we can see from Eq.~(\ref{eq:zeta_total}), $\zeta$ has inflationary and curvaton contributions. We follow the usual practice of considering the scenario where the curvaton dominates the curvature perturbation, that is, $r_{D}\zeta_\sigma\gg \left(1-r_{D}\right)\zeta_{\phi}$, and so we may use the approximation \cite{Moroi:2002rd,Lyth:2003ip}
\begin{equation}
\zeta\simeq r_{D}\zeta_{\sigma}.
\end{equation} 

This also allows us to assume that the spectral index of all perturbation spectra (adiabatic and isocurvature) is given by one value $n_{s}$. 

The curvaton is a massive scalar field, and so for the simplest quadratic curvaton potentials, the curvaton energy density is $\rho_{\sigma}\sim \overline{\sigma}^{2}+2\overline{\sigma}\delta\sigma+\left(\delta\sigma\right)^{2}$, where $\overline{\sigma}$ is the homogeneous value of $\sigma$ and $\delta\sigma$ a spatial perturbation. As $\sigma$ itself is a Gaussian random field, $\rho_\sigma$ is non-Gaussian. The resulting non-Gaussianity is of local type, that is, $\zeta=\zeta_{g}(\vec{x})+\frac{3}{5}f_{\rm nl}\left[\zeta_g^{2}(\vec{x})-\langle \zeta_{g}^{2}(\vec{x})\rangle\right]$with \begin{equation}f_{\rm nl}=\frac{5}{4r_{D}}-\frac{5r_{D}}{6}-\frac{5}{3},\label{eq:fnl_predict}\end{equation} where $\zeta_{g}(\vec{x})$ is a Gaussian random field \cite{Lyth:2001nq,Lyth:2002my,Gordon:2002gv,Malik:2006pm,Kawasaki:2011pd,Ade:2015lrj}. The stringent limits to local-type non-Gaussianity from \textit{Planck} temperature data, $f_{\rm nl}=2.7\pm 5.8$ impose the constraint $r_{D}>0.12$ \cite{Ade:2015lrj,Ade:2015ava}. These constraints do not depend on curvaton-decay scenario, and are thus relatively model-independent. In some curvaton-decay scenarios, residual isocurvature perturbations would be excited, making more stringent limits to $r_{D}$ possible. Additionally, limits to or a detection of curvaton-type isocurvature would make it possible to test the decay physics of the curvaton.

If the densities of all species are determined after curvaton decay, then the density perturbations in all species are set by $\zeta$ alone, leading to purely adiabatic fluctuations. On the other hand, if some conserved quantum numbers are generated by or before curvaton decay while others are not, there is a mismatch in density fluctuations, leading to a gauge-invariant entropy (or isocurvature) perturbation. In particular \cite{Lyth:2001nq,Lyth:2002my,Gordon:2002gv},
\begin{eqnarray}
\tilde{\zeta}_{x}=\left\{\begin{array}{ll}0,&\mbox{if $x$ is produced before $\sigma$ decay,}\\\tilde{\zeta}_{\sigma},&\mbox{if $x$ is produced by $\sigma$ decay},\\ \zeta,&\mbox{if $x$ is produced after $\sigma$ decay}.\end{array}\label{eq:history}\right.
\end{eqnarray} 

Here the index $x$ denotes $b$ (baryon number), $L$ (lepton number), or $c$ (CDM). The $\tilde{\zeta}_{x}$ indicates initial curvature fluctuations on hypersurfaces of constant particle $\textit{number}$ (for CDM) or conserved quantum number (in the case of baryons or leptons). The curvaton is assumed to behave as matter at the relevant epochs, and so $\tilde{\zeta}_{\sigma}=\zeta_{\sigma}$.

We distinguish between quantum numbers (like baryon and lepton number) and densities, as baryon and lepton number could be generated at very early times, long before quarks bind to produce actual baryons. Indeed, baryogenesis (which refers to the creation of baryon \textit{number}) could be related to curvaton physics, even if the production of actual baryons happens much later.

The gauge-invariant entropy fluctuation between $x$ and photons is given by
\begin{equation}
S_{x \gamma}\equiv 3(\zeta_{x}-\zeta_{\gamma})
\end{equation}and is conserved on super-horizon scales \cite{Bardeen:1980kt,Mukhanov:1990me,Malik:2004tf}, as long as the equation of state of the species $i$ (or the carriers of the relevant quantum number) is constant and the quantum numbers are conserved. Photon perturbations are described by $\zeta_{\gamma}$, the spatial curvature perturbation on hyper-surfaces of constant photon energy-density. For baryons or leptons $\zeta_{x}$ is the curvature perturbation on surfaces of constant energy density of whichever species carries the quantum number (at late times, these would be actual surfaces of constant baryon energy density). 

The constant super-horizon values of $\zeta$ and $S_{x\gamma}$ are `initial conditions' which precede horizon entry and determine the spectra of CMB anisotropies, as computed by \textsc{camb} \cite{cambnotes} or any other CMB Boltzmann code. 
We take the initial values $S_{x\gamma}$ to be defined at some time after the relevant species thermally decouple and reach their final equation of state (for example, if $x=c$, we consider $S_{c\gamma}$ at some time after CDM has become nonrelativistic). After the quantum number associated with $x$ thermally freezes out, $\tilde{\zeta}_{x}$ is conserved on super-horizon scales because the relevant quantum numbers are conserved. If $x\in \left\{c,b\right\}$, $S_{x\gamma}$ is set long after actual baryons and CDM become non-relativistic, and so $\zeta_{x}=\tilde{\zeta}_{x}$, because surfaces of constant energy and number density coincide. We discuss the subtler case of lepton number fluctuations and neutrino isocurvature in Sec. \ref{sec:decay_scen}. 

For any quantum number/species, there are then $3$ scenarios \cite{Lyth:2003ip}:
\begin{widetext}
\begin{eqnarray}
S_{x\gamma}=\left\{\begin{array}{ll}-3\zeta-3(\zeta_{\gamma}-\zeta),&\mbox{if $x$ is produced before $\sigma$ decay,}\\3\left(\frac{1}{r_{D}}-1\right)\zeta-3(\zeta_{\gamma}-\zeta),&\mbox{if $x$ is produced by $\sigma$ decay},\\ -3(\zeta_\gamma-\zeta),&\mbox{if $x$ is produced after $\sigma$ decay}.\end{array}\right.\label{eq:strew}
\end{eqnarray} \end{widetext} When fluctuations are set by the curvaton, as we can see from Eq.~(\ref{eq:strew}), entropy fluctuations are set completely by the adiabatic fluctuation (as we would expect when only fluctuations in a single field are important), and are thus totally correlated or anti-correlated to $\zeta$. 

Anti-correlated isocurvature perturbations can lower the observed CMB temperature anisotropy at low multipole $l$, improving the mild observed tension between the best-fit $\Lambda$CDM model and large-scale CMB observations \cite{Hinshaw:2012aka,Ade:2013zuv}. To see what this fact implies for curvaton physics, and to more broadly test the curvaton model using CMB observations, we now derive the isocurvature amplitudes in different curvaton-decay scenarios. To simplify the discussion, we will describe curvaton-decay scenarios with the notation $(b_{y_{b}},c_{y_{c}},L_{y_{L}})$, where $y_{L}\in \left\{{\rm before,by,after}\right\}$. For example, $(b_{\rm by},c_{\rm after}, L_{\rm before})$ indicates a curvaton-decay scenario in which baryon number is generated by curvaton decay, cold dark-matter after curvaton decay, and lepton number before curvaton decay.

\section{Curvaton-decay scenarios}
\label{sec:decay_scen}
The various curvaton-decay scenarios can be divided into cases where the production of either the baryon number, lepton number, or CDM occurs before the curvaton decays, by the curvaton decay, or after the curvaton decays. This naturally leads to a total of $3\times 3 \times 3= 27$ distinct scenarios.  As discussed in the previous Section, curvaton decay can occur at any time after inflation ends. Curvaton decay must certainly also occur before big-bang nucleosynthesis (BBN).  This means that within the single-field slow-roll inflationary models the curvaton may decay at temperatures ranging between $10^{16}$ GeV \cite{Ade:2015lrj} down to $\sim 4$ MeV \cite{Hannestad:2004px}, at which point the primordial light elements must be produced.  In order for all 27 scenarios to be realized, there must be mechanisms that generate baryon number, lepton number, and CDM over this wide range of energy scales, as we now discuss.

A persistent mystery is the origin of baryon number-- i.e., the observed net asymmetry of baryons over anti-baryons in the universe. Plausible models bracket a range of energy scales, from baryogenesis at the electroweak scale \cite{Morrissey:2012db} to direct production of baryon number through a coupling to the inflaton or curvaton (see Ref.~\cite{Lemoine:2006sc} and references therein). The energy scale of baryogenesis could thus be anywhere in the range $1~{\rm TeV}$--$10^{16}~{\rm GeV}$. Since both the inflationary energy scale and the energy scale of curvaton relevance/decay are poorly constrained, it is possible for baryon number to be produced before, by, or after curvaton decay. 

The observed baryon asymmetry could be produced through partial conversion of a much larger primordial lepton asymmetry. One of the ways (reviewed at length in Ref. \cite{Davidson:2008bu}) to account for the observed non-zero neutrino mass is to invoke the seesaw mechanism \cite{Minkowski:1977sc,Mohapatra:1980yp}.  The seesaw mechanism generically introduces a hierarchy of neutrinos with masses above the electroweak scale leading to the generation of lepton number at temperatures greater than $\sim 100$ GeV.  

Alternatively, lepton number could be produced near the end of inflation (at energies as high as $\sim 10^{16}$ GeV), perhaps by a Chern-Simons (parity-violating) terms in the gravitational sector \cite{Alexander:2004us} or by a novel coupling of chiral fermions to an axion-like field \cite{Adshead:2015jza}. On the other hand the $\nu$MSM model \cite{Asaka:2005an} allows for lepton number to be generated at lower energies.  Finally, as discussed in the previous Section, it is possible that the decay of the curvaton field produces lepton number, leading to isocurvature perturbations in the neutrino density perturbations. 

The identity and production mechanism of the CDM is also a mystery \cite{Zwicky:1933gu,Rubin:1970zza,Rubin:1980zd,Feng:2010gw}. One possibility is that the CDM consists of weakly interacting massive particles (WIMPs) thermally produced by physics at the $\sim$ TeV scale \cite{Jungman:1995df}.  If this is so, the CDM would be produced around the electroweak energy scale. Direct-detection experiments, however, have placed increasingly stringent limits to WIMP couplings. The most natural WIMP candidate (a stable super-partner in supersymmetric models) is also under increasing pressure from experiment, due to the lack of evidence for low-energy supersymmetry from the Large Hadron Collider (LHC) (see Ref.~\cite{Catena:2013pka} and references therein). This motivates the consideration of other CDM candidates.

One possible CDM candidate is a stable extremely massive particle (or \textit{wimpzilla}) with mass in the range $10^{12}~{\rm GeV}\lsim m\lsim 10^{16}~{\rm GeV}$ \cite{Chung:1998zb}. The wimpzilla might be produced by gravitational particle production during inflation or directly from inflaton decay \cite{Kofman:1994rk,Chung:1998zb}. Similarly, even a standard lighter supersymmetric WIMP could be produced by curvaton decay if WIMPs couple to the curvaton field \cite{Lyth:2002my}. Just as with baryon and lepton numbers, CDM could thus be produced before, by, or after curvaton decay \cite{Gordon:2002gv}. 

Altogether, there are a variety of logically possible scenarios for producing the correlated isocurvature fluctuations discussed in Sec. \ref{sec:curv_model}. Our goal in this work is to test these scenarios using CMB data. We assemble for the first time in one work expressions for the amplitude of correlated baryon, CDM, and neutrino isocurvature-density (NID) perturbations in all $27$ possible curvaton-decay scenarios, as shown in Table \ref{tab:curvatoncases} and Eq.~(\ref{eq:finalneutrino}).  

This allows us to build on past work, which explored only one isocurvature mode at a time \cite{DiValentino:2011sv} or neglected NID perturbations \cite{Gordon:2002gv,He:2015msa}, and self-consistently test for the first time the full parameter space of $r_{D} $ and $\xi_{\rm lep}$ in all $27$ curvaton-decay scenarios. At the level of observable power spectra in linear perturbation theory, the CDM and baryon isocurvature modes are indistinguishable \cite{Gordon:2002gv,cambnotes,Grin:2011tf,Grin:2013uya,He:2015msa}, but the NID mode has a distinct physical imprint \cite{DiValentino:2011sv,Kasanda:2011np} from the others that can be separately probed using the data.

We begin with the simplest curvaton-decay scenarios, in which there is no lepton asymmetry $L=\Delta n_{L}/n_{L}$ (here  $\Delta n_{L}=n_{L}-n_{\overline{L}}$, where $n_{L}$ and $n_{\overline{L}}$ denote the number densities of lepton number and anti-lepton number, respectively). 

\subsection{No lepton asymmetry}
\label{sec:nolepton}
During radiation domination, the total curvature perturbation is given by
\begin{equation}
\zeta=(1-R_{\nu})\zeta_{\gamma}+R_{\nu}\zeta_{\nu},\label{eq:ztot}
\end{equation}where $R_{\nu}\equiv \rho_{\nu}/(\rho_{\gamma}+\rho_{\nu})$ is the energy fraction in massless neutrinos, a constant after electron-positron annihilation. Neutrinos carry lepton number and thermally decouple near temperatures $T\sim 2~{\rm MeV}$. If there is no lepton asymmetry, spatial fluctuations in lepton number density track the total energy density, and so $\zeta_{\nu}=\zeta$. From Eq.~(\ref{eq:ztot}) we then see that $\zeta=\zeta_{\gamma}=\zeta_{\nu}$ and thus $S_{\nu \gamma}=0$. It is then straightforward to obtain the relationships between baryon/CDM entropy fluctations and curvature fluctuations for a variety of curvaton-decay scenarios by applying Eq.~(\ref{eq:strew}). The resulting amplitudes are shown in Table \ref{tab:curvatoncases}. Later, to interpret constraints, it is useful to define the total isocurvature in non-relativistic matter:\begin{align}
S_{m \gamma}\equiv &3 \left[\left(\frac{\Omega_{b}}{\Omega_{m}}\zeta_{b}+\frac{\Omega_{c}}{\Omega_{m}}\zeta_{c}\right)-\zeta\right]\nonumber \\=&\left[\frac{\Omega_{b}}{\Omega_{m}}S_{b\gamma}+\frac{\Omega_{c}}{\Omega_{m}}S_{c\gamma}\right].
\end{align}Here $\Omega_{b}$ and $\Omega_{c}$ are the usual relic densities of baryons and CDM relative to the cosmological critical density.

We note that the scenarios  $(b_{\rm by},c_{\rm before},L_{y_L})$ and $(b_{\rm before},c_{\rm by},L_{y_L})$ lead to correlated (or anti-correlated) isocurvature perturbations. These scenarios mitigate some of the tension between CMB data (for $l\lsim 50$) and the best-fit $\Lambda$CDM model \cite{Hinshaw:2012aka,Ade:2013zuv,Ade:2015xua}. We discuss this further in Sec \ref{sec:data}. 

The near cancellation of baryon and CDM isocurvature contributions to $S_{m\gamma}$ in these scenarios requires fine-tuned values of $r_{D}\sim \Omega_{b}/\Omega_{m}$ and $r_{D}\sim \Omega_{c}/\Omega_{m}$. This yields a relatively large CIP amplitude of $S_{bc}=3\left(\zeta_{b}-\zeta_{c}\right)=3\zeta/r_{\rm D}$ and $-3\zeta/r_{\rm D}$ in the $(b_{\rm by},c_{\rm before},L_/{y_{L}})$ and $(b_{\rm before},c_{\rm by},L_{y_{L}})$ scenarios, respectively, or more explicitly, $S_{bc}\simeq 20 \zeta$ and $S_{bc}\simeq 3.5\zeta$. These CIP amplitudes could leave observable imprints on off-diagonal correlations (or equivalently, the CMB bispectrum and trispectrum), a possibility discussed further in Ref. \cite{Grin:2011tf,Grin:2013uya}. The cases $(b_{\rm before},c_{\rm after},L_{y_{L}})$, $(b_{\rm after},c_{\rm before},L_{y_{L}})$, and $(b_{\rm before},c_{\rm before},L_{y_{L}})$ are completely ruled out by the data, as already shown in Refs. \cite{Gordon:2002gv}. We do not consider them further.

The situation is considerably richer if there is a net lepton asymmetry. As we see in Sec. \ref{sec:lep}, if the lepton symmetry is generated before or after $\sigma$ decay, the ratios $S_{b\gamma}/\zeta,S_{c\gamma}/\zeta$, and $S_{m \gamma}/\zeta$ are given (to very good or perfect accuracy, respectively) by the values shown in Table \ref{tab:curvatoncases} with $S_{\nu \gamma}$=0. On the other hand, if the lepton asymmetry is generated by $\sigma$ decay, there is a residual neutrino isocurvature perturbation $S_{\nu\gamma}$ \cite{Lyth:2001nq,Lyth:2003ip,Gordon:2003hw}.
\subsection{Lepton asymmetry}
\label{sec:lep}
Each neutrino species carries the lepton number of the corresponding lepton flavor, and so in the presence of a lepton asymmetry, fluctuations in $\Delta n_{L}$ result in neutrino isocurvature perturbations. For massless neutrinos, the occupation number is
\begin{equation}
f_{j}(E)=\left[e^{E/T_{\nu}\mp \xi_{j}}+1\right]^{-1},\end{equation}
where the flavor label takes values $j={\rm e},\mu,$ or $\tau$, the corresponding chemical potential $\xi_{j}$ parameterizes the primordial lepton asymmetry, the minus sign applies for neutrinos, and the plus sign applies for anti-neutrinos. Unlike the cosmological baryon asymmetry $\eta \simeq 6\times 10^{-10}$, $\xi_{j}$ is rather poorly constrained. Some models of baryogenesis require comparable levels of lepton and baryon asymmetry, but others convert a much larger lepton asymmetry into the experimentally known baryon asymmetry. Electron neutrinos (whose number density depends on $\xi_{e^{-}}$) set the rates of $\beta$-decay processes active during BBN, and so the value of $\xi_{e^{-}}$ effects the primordial neutron-to-proton ratio $n/p\propto {\rm exp}(-\xi_{e^{-}})$ and the resulting abundance of $\lsup{\rm{He}}{4}$ \cite{Dolgov:2002ab,Cuoco:2003cu}. 

A lepton asymmetry also alters $N_{\rm eff}$, the number of relativistic degrees of freedom during BBN, although this effect is less important for setting abundances than the altered $n/p$ ratio. Neutrinos are now known to have mass and as a result exhibit flavor oscillations. Independent of initial conditions, solar neutrino observations and results from the KamLAND experiment \cite{Dolgov:2002ab,Guzzo:2003xk} indicate $\nu$ mass splittings and mixing angles that would lead to near flavor equilibrium early on, and so $\xi_{e^{-}}=\xi_{\mu}=\xi_{\tau}\equiv\xi_{\rm lep}$. BBN abundances (including the $\lsup{\rm{He}}{4}$ abundance $Y_{\rm He}$) depend not only on the primordial values $\xi_{\rm lep}$, but also on the mixing angles between neutrinos \cite{Dolgov:2002ab}, and in particular on the value of $\theta_{13}$. Current reactor and long-baseline neutrino experiments indicate that $\sin^{2}(\theta_{13})\simeq 0.03$, giving a $95\%$-confidence BBN limit of $\xi_{\rm lep} \leq 0.03$ \cite{Mangano:2011ip}.
\begin{figure}[h!]
\resizebox{!}{7.5cm}{\includegraphics{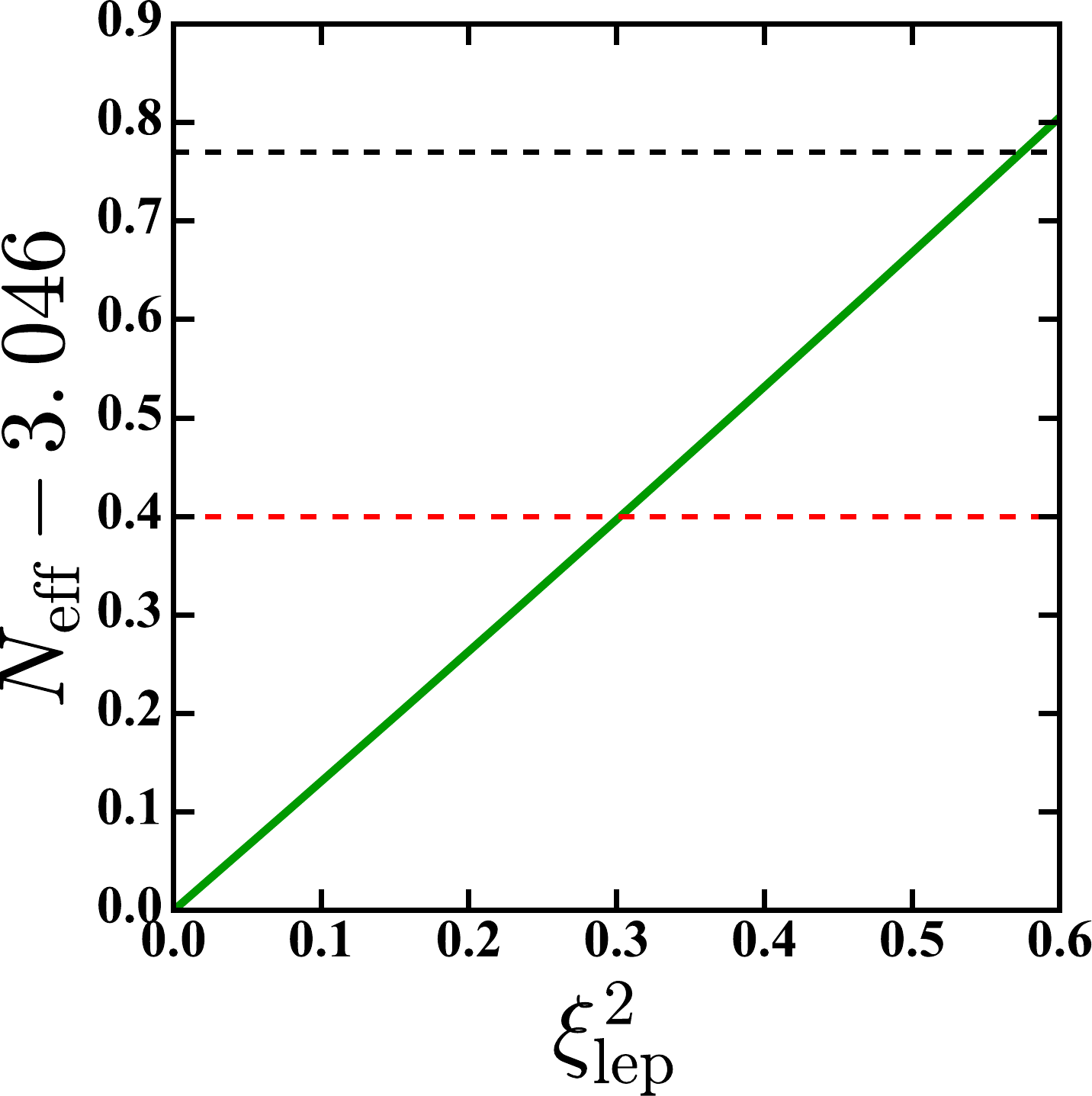}}
\caption{The relationship between $N_{\rm eff}$ and $\xi_{\rm lep}^2$ [Eq~(\ref{eq:neff})].  The black dashed-line indicates the 95\% CL upper limit to $\Delta N_{\rm eff}$ from the 2015 \textit{Planck} analysis using TT+LowP+BAO  and corresponds to a 95\% CL upper limit $\xi_{\rm lep}^2 \leqslant 0.5$; the red dashed-line indicates the 95\% CL upper limit using TT+AllP+BAO.}
\label{fig:Neff}
\end{figure} 

The resulting $\nu$ energy and lepton- number densities are \cite{Lyth:2002my,Gordon:2003hw,DiValentino:2011sv}
\begin{align}
\frac{\rho_{i}}{\rho_{\gamma}}=&\frac{7}{8}\left(\frac{T_{\nu}}{T_{\gamma}}\right)^{4}A_{i},\\\frac{\Delta n_{i}}{n_{\gamma}}=&2.15 \left(\frac{T_{\nu}}{T_\gamma}\right)^{3}B_{i},\\
A_{i}=&\left[1+\frac{30}{7}\left(\frac{\xi_{\rm lep}}{\pi}\right)^{2}+\frac{15}{7}\left(\frac{\xi_{\rm lep}}{\pi}\right)^{4}\right],\\
B_{i}=&\left[\left(\frac{\xi_{\rm lep}}{\pi}\right)^{2}+\left(\frac{\xi_{\rm lep}}{\pi}\right)^{4}\right],
\end{align}
which can also be parameterized as \cite{Lyth:2002my}
\begin{equation}
N_{\rm eff}=3.046+3\left[\frac{30}{7}\left(\frac{\xi_{\rm lep}}{\pi}\right)^{2}+\frac{15}{7}\left(\frac{\xi_{\rm lep}}{\pi}\right)^{4}\right].
\label{eq:neff}
\end{equation}

Past forecasts and recent analyses of \textit{Planck} data show that if the only effects of $\xi_{\rm lep}$ are to alter $N_{\rm eff}$ and the free-electron fraction (by altering $Y_{\rm He}$) at decoupling, CMB constraints to $\xi_{\rm lep}$ (shown in Fig.~\ref{fig:Neff}) will remain less sensitive than constraints from astronomical measurements of primordial element abundances \cite{Kinney:1999pd,Bowen:2001in,Popa:2008tb,Shiraishi:2009fu,Caramete:2013bua,Grohs:2014rea,Grohs:2015eua}. In the curvaton scenario, however, if the lepton asymmetry is generated by curvaton decay, the amplitude of neutrino-isocurvature-density fluctuations depends on the values of $\xi_{\rm lep}$, offering an additional possible channel for constraining this parameter. Neutrino experiments may still yield surprises as to the precise values of quantities like $\theta_{13}$. We thus explore what constraints to the neutrino sector are possible from CMB observations alone. In the future, measurements of the 21-cm emission/absorption power spectrum from neutral hydrogen (during the epoch of reionization or during the cosmic dark ages) could be useful probes of the value of $\xi_{\rm lep}$ \cite{Kohri:2014hea}.

We assume that the cosmic thermal history is conventional between neutrino decoupling and electron-positron annihilation, and thus neglect fluctuations in the neutrino-photon temperature ratio $T_{\nu}/T_{\gamma}\simeq (4/11)^{1/3}$. It is then straightforward to show that for neutrinos \cite{Gordon:2003hw}
\begin{align}
\zeta_i-\zeta_\gamma=&\frac{1}{4}\frac{A'_{i}}{A_{i}}\frac{\delta \xi_{\rm lep}}{\pi},\\
\tilde{\zeta}_i-\zeta_\gamma=&\frac{1}{3}\frac{B'_{i}}{B_{i}}\frac{\delta \xi_{\rm lep}}{\pi}.
\end{align}Neutrinos inherit the lepton asymmetry and its fluctuations, and so $\delta \xi_{\rm lep}=\pi B_{i} \tilde{S}_{L}/B'_{i}$, where $\tilde{S}_{L}=3(\tilde{\zeta}_{L}-{\zeta}_{\gamma})$. We then see that\begin{equation}
S_{\nu\gamma}\simeq\frac{15}{7}\sum_{i=\mu,e,\tau}\left(\frac{\xi_{\rm lep}}{\pi}\right)^{2}\tilde{S}_{L},
\end{equation}where we have assumed that $\xi_{\rm lep}/\pi\ll 1$ and assumed that flavor mixing of the cosmic neutrino background is negligible after neutrino decoupling.  We thus have that \cite{Gordon:2003hw,DiValentino:2011sv}
\begin{equation}
S_{\nu \gamma}\simeq\frac{45}{7}\left(\frac{\xi^{2}_{\rm lep}}{\pi^{2}}\right)\tilde{S}_{L}=\frac{135}{7}\left(\frac{\xi_{\rm lep}^{2}}{\pi^{2}}\right)\left(\tilde{\zeta}_{L}-{\zeta}_\gamma\right).
\end{equation}
To proceed further, we must specify when lepton number (L) is generated. Applying Eq.~(\ref{eq:history}), we obtain \cite{Gordon:2003hw,DiValentino:2011sv} \begin{widetext}
\begin{align}
S_{\nu\gamma}=\left\{\begin{array}{ll}-\frac{135}{7}\left(\frac{\xi_{\rm lep}}{\pi}\right)^2\zeta_{\gamma}&\mbox{if $L$ is generated before $\sigma$ decay},\\ 
\frac{135}{7}\left(\frac{\xi_{\rm lep}}{\pi}\right)^2\left(\frac{\zeta}{r_{\rm D}}-\zeta_\gamma\right)&\mbox{if $L$ is generated by $\sigma$ decay},\\
\frac{135}{7}\left(\frac{\xi_{\rm lep}}{\pi}\right)^2\left(\zeta-\zeta_{\gamma}\right)&\mbox{if $L$ is generated after $\sigma$ decay.}
\end{array}\right.\label{eq:niso_lep}
\end{align}
\end{widetext}
Substituting into Eq.~(\ref{eq:ztot}) and solving for $S_{\nu\gamma}$, we then obtain [to lowest order in $(\xi_{\rm lep}/\pi)^{2}$] \cite{Gordon:2003hw}:
\begin{widetext}
\begin{align}
\frac{S_{\nu\gamma}}{\zeta}=\left\{
\begin{array}{ll}
0&\mbox{if $L$ is generated before $\sigma$ decay},\\
\frac{135}{7}\left(\frac{\xi_{\rm lep}}{\pi}\right)^2\left(\frac{1}{r_{\rm D}}-1\right)&\mbox{if $L$ is generated by $\sigma$ decay},\\
0&\mbox{if $L$ is generated after $\sigma$ decay.}
\end{array}
\label{eq:finalneutrino}
\right.
\end{align}
\end{widetext}
The expression for the case of $L$ generated before $\sigma$ decay is approximate, and has corrections of order $S_{\nu\gamma}\sim 10^{-2}\zeta$ which are negligible at the level of accuracy needed for the MCMC analysis of Sec. \ref{sec:data}. The expression for the case of $L$ generated after $\sigma$ decay results from the requirement that the penultimate equation hold independent of the true values of $R_{\nu}$ and $\xi_{\rm lep}$.

 In scenarios where $L$ is generated by $\sigma$ decay, there is a mismatch between the total $\zeta$ (which has contributions from neutrinos and photons) and $\zeta_{\gamma}$. This must be self-consistently included in Eq.~(\ref{eq:ztot}) to obtain the correct expressions for the relationships between $S_{b\gamma}$ (or $S_{c\gamma}$) and $\zeta$, shown in Table \ref{tab:curvatoncases}. If lepton number is generated before or after $\sigma$ decay, the amplitudes are given as before in Table \ref{tab:curvatoncases} with $S_{\nu \gamma}=0$. 
  
  \begin{table*}[hts]
\caption{Baryon and CDM isocurvature amplitudes (in terms of the curvature perturbation $\zeta$) for the various curvaton-decay scenarios. If the lepton chemical potential $\xi_{\rm lep}=0$, $S_{\nu \gamma}=0$. Otherwise, if $\xi_{\rm lep}\neq 0$ \textit{and} there is a net lepton number $L\neq 0$, $S_{\nu \gamma}$ is given by Eq.~(\ref{eq:niso_lep}), taking non-zero values only if $L$ is generated by curvaton decay, that is, $y_{L}={\rm by}$.
This is discussed in detail in Secs. \ref{sec:nolepton} and \ref{sec:lep}. The notation $(b_{y_{b}},c_{y_{c}},L_{y_{L}})$ for various curvaton-decay scenarios is introduced in Sec. \ref{sec:curv_model}.}
\label{tab:curvatoncases}
\begin{ruledtabular}
\begin{tabular}{c|ccc}
\rule{0pt}{4.5ex}
\rule[-3ex]{0pt}{0pt}
scenario& $\dfrac{S_{b\gamma}}{\zeta}$ 
& $\dfrac{S_{c\gamma}}{\zeta}$ &$\dfrac{S_{m\gamma}}{\zeta}$\\ \hline
\rule{0pt}{4.5ex}
\rule[-3ex]{0pt}{0pt}
$(b_{\rm by},c_{\rm before},L_{y_{L}})$
& $3\left(\dfrac{1}{r_{D}}-1\right)+R_{\nu}\dfrac{S_{\nu\gamma}}{\zeta}$ 
& $-3+R_{\nu}\dfrac{S_{\nu\gamma}}{\zeta}$ &$3\left(\dfrac{\Omega_{b}}{\Omega_{m}r_{D}}-1\right)+R_{\nu}\dfrac{S_{\nu\gamma}}{\zeta}$

\\
\rule{0pt}{4.5ex}
\rule[-3ex]{0pt}{0pt}
$(b_{\rm before},c_{\rm by},L_{y_{L}})$
& $-3+R_{\nu}\dfrac{S_{\nu\gamma}}{\zeta}$
& $3\left(\dfrac{1}{r_{D}}-1\right)+R_{\nu}\dfrac{S_{\nu\gamma}}{\zeta}$&$3\left(\dfrac{\Omega_{c}}{\Omega_{m}r_{D}}-1\right)+R_{\nu}\dfrac{S_{\nu \gamma}}{\zeta}$

\\
\rule{0pt}{4.5ex}
\rule[-3ex]{0pt}{0pt}
$(b_{\rm by},c_{\rm after},L_{y_{L}})$
&  $3 \left(\dfrac{1}{r_{D}}-1\right)+R_{\nu}\dfrac{S_{\nu\gamma}}{\zeta}$
&  $R_{\nu}\dfrac{S_{\nu\gamma}}{\zeta}$&$3\dfrac{\Omega_{b}}{\Omega_{m}}\left(\dfrac{1}{r_{D}}-1\right)+R_{\nu}\dfrac{S_{\nu\gamma}}{\zeta}$

\\
\rule{0pt}{4.5ex}
\rule[-3ex]{0pt}{0pt}
$(b_{\rm after},c_{\rm by},L_{y_{L}})$
&  $R_{\nu}\dfrac{S_{\nu\gamma}}{\zeta}$
& $3\left(\dfrac{1}{r_{\rm D}}-1\right)+\dfrac{S_{b\gamma}}{\zeta}$&$3\dfrac{\Omega_{c}}{\Omega_{m}}\left(\dfrac{1}{r_{D}}-1\right)+R_{\nu}\dfrac{S_{\nu\gamma}}{\zeta}$

\\
\rule{0pt}{4.5ex}
\rule[-3ex]{0pt}{0pt}
$(b_{\rm before},c_{\rm after},L_{y_{L}})$
& $-3+R_{\nu}\dfrac{S_{\nu\gamma}}{\zeta}$
& $R_{\nu}\dfrac{S_{\nu\gamma}}{\zeta}$&$-3\dfrac{\Omega_{b}}{\Omega_{m}}+R_{\nu}\dfrac{S_{\nu\gamma}}{\zeta}$

\\
\rule{0pt}{4.5ex}
\rule[-3ex]{0pt}{0pt}
$(b_{\rm after},c_{\rm before},L_{y_{L}})$
& $ R_{\nu}\dfrac{S_{\nu\gamma}}{\zeta}$ 
&$-3+\dfrac{S_{b\gamma}}{\zeta}$ &$-3\dfrac{\Omega_{c}}{\Omega_{m}}+R_{\nu}\dfrac{S_{\nu\gamma}}{\zeta}$

\\
\rule{0pt}{4.5ex}
\rule[-3ex]{0pt}{0pt}
$(b_{\rm before},c_{\rm before},L_{y_{L}})$
&  $-3+R_{\nu}\dfrac{S_{\nu \gamma}}{\zeta}$
&  $\dfrac{S_{b\gamma}}{\zeta}$&$\dfrac{S_{b\gamma}}{\zeta}$

\\
\rule{0pt}{4.5ex}
\rule[-3ex]{0pt}{0pt}
$(b_{\rm by},c_{\rm by},L_{y_{L}})$
&$3\left(\dfrac{1}{r_{D}}-1\right) +R_{\nu}\dfrac{S_{\nu \gamma}}{\zeta}$
& $\dfrac{S_{b \gamma}}{\zeta}$& $\dfrac{S_{b\gamma}}{\zeta}$

\rule{0pt}{4.5ex}
\rule[-3ex]{0pt}{0pt}
\\
$(b_{\rm after},c_{\rm after},L_{y_{L}})$
& $R_{\nu}\dfrac{S_{\nu\gamma}}{\zeta}$
& $R_{\nu}\dfrac{S_{\nu\gamma}}{\zeta}$&$R_{\nu}\dfrac{S_{\nu\gamma}}{\zeta}$
\rule{0pt}{4.5ex}
\rule[-3ex]{0pt}{0pt}
\\
\end{tabular}
\end{ruledtabular}
\end{table*}

\section{Data}
\label{sec:data}
The main effect of the curvaton model is to introduce totally correlated (or anti-correlated) isocurvature modes into the initial conditions of the cosmological perturbations.  In order to test the various curvaton decay channels we use the CMB temperature and E-mode polarization power-spectra measured by the Planck satellite \cite{Ade:2013kta,Ade:2015xua,Aghanim:2015xee}.  The large-scale E-mode measurements mainly constrain the optical depth to the surface of last scattering, $\tau$, while the small-scale E-mode measurements provide additional constraints on the allowed level of isocurvature \cite{Bucher:2000hy}. We also use measurements of Baryon Acoustic Oscillations (BAOs) \cite{Anderson:2013zyy,Ross:2014qpa} to break geometric degeneracies in the CMB data and thus improve the sensitivity of the \textit{Planck} data to isocurvature perturbations. 

\begin{figure*}
\begin{center}
\resizebox{!}{11cm}{\includegraphics{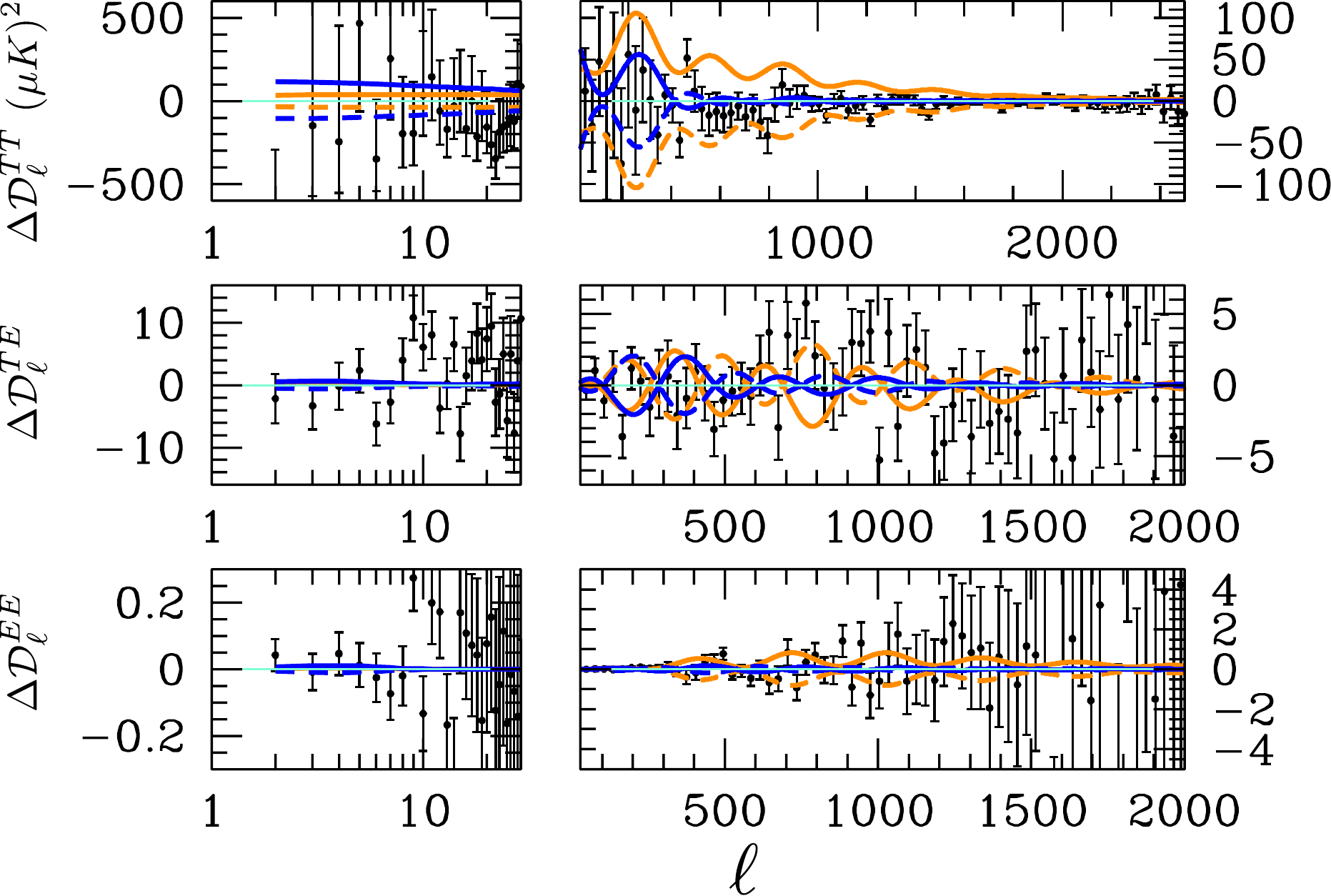}}
\caption{A comparison between the difference between a purely adiabatic mode and a totally correlated (solid) or anticorrelated (dashed) matter (blue) or neutrino density (orange) isocurvature mode.  Each panel shows the binned residuals $\Delta \mathcal{D}^{XY}_{\ell} \equiv \ell(\ell+1) \Delta C^{XY}_{\ell}/(2\pi)$ (see Ref.~\cite{Aghanim:2015xee} for details on the binning procedure).  The matter isocurvature has an amplitude $S_{m \gamma} = 0.2$ and the neutrino density isocurvature has an amplitude $S_{\nu \gamma} = 0.1$. We also show the residuals for the power spectrum measured by the \textit{Planck} satellite \cite{Aghanim:2015xee}. Note that the horizontal scale is logarithmic up to $\ell = 29$ and then is linear; the vertical scale on the left and right-hand sides are different.}
\label{fig:comp1}
\end{center}
\end{figure*}

The introduction of matter isocurvature modes, shown by the blue curves in Fig.~\ref{fig:comp1}, has its most significant affect on the large-scale TT and TE power spectrum, where it changes the height of the Sachs-Wolfe plateau and alters the shape/amplitude of the Integrated Sachs-Wolfe (ISW) effect. On the other hand, neutrino-density isocurvature with a comparable amplitude, shown by the orange curves in Fig.~\ref{fig:comp1}, effects CMB anisotropies more dramatically at all scales. 

The \textit{Planck} data has been divided up into a large angular-scale dataset (low multipole number) and a small angular-scale dataset (high multipole number) \cite{Aghanim:2015xee}.  For all constraints we use the entire range of measurements for the TT power spectrum as well as the low multipole polarization (TE and EE) data, which we denote as LowP.  We also compute constraints using the entire multipole range of polarization measurements, denoted by AllP. The division between these two datasets is the multipole number $\ell =29$ which approximately corresponds to an angular scale of $\simeq 5^\circ$. 

As demonstrated in Fig.~\ref{fig:comp1} polarization data can break degeneracies present in a temperature-only analysis.  This statement is especially true for tests of the adiabaticity of the initial conditions \cite{Bucher:2000hy}.   The analysis in Sec.~11 of Ref.~\cite{Ade:2015lrj} and Sec.~6.2.3 of Ref.~\cite{Ade:2015xua} includes constraints to isocurvature modes using the \textit{Planck} 2015 data.  As they point out the addition of AllP greatly improves the constraint to isocurvature modes which are correlated to the adiabatic mode.  

For example, the fractional contribution to the temperature power spectrum is constrained to $\alpha = -0.0025^{+ 0.0035}_{- 0.0047}$ at 95\% CL using \textit{Planck} TT + LowP where the sign of $\alpha$ indicates whether the isocurvature contribution is totally correlated ($\alpha >0$) or anti-correlated ($\alpha <0$) with the adiabatic mode.  The preference 
for an anitcorrelated mode comes from the well-known deficit of power on large angular scales \cite{Ade:2013kta,Schwarz:2015cma}. 
When all polarization data is included in the analysis the centroid shifts upward and the overall uncertainty on $\alpha$ is reduced by more than 50\%: $\alpha = 0.0003^{+ 0.0016}_{-0.0012}$ at 95\% CL. As noted by the \textit{Planck} team \cite{Ade:2015xua} these effects may both be driven by a significantly low point in the TE cross power spectrum which may be due to unidentified systematic effects (see, e.g., Ref.~\cite{Addison:2015wyg}). 

In order to highlight the effects of including all of the publicly available \textit{Planck} data we divide our analysis into two sets of data: \textit{Planck} TT+BAO+LowP and \textit{Planck} TT+BAO+AllP.  Given the uncertainty around systematic effects in the high-$\ell$ polarization power spectrum we take the \textit{Planck} TT+BAO+LowP constraints to be more robust. 

In order to compare the data to our model we use a modified version of the publicly available Boltzmann code \textsc{CosmoMC} \cite{Lewis:2002ah} along with the publicly available Planck Likelihood code \cite{Aghanim:2015xee} included with the 2015 Planck data release.  We made modifications to these codes in order to include the two curvaton parameters $r_D$ and $\xi_{\rm lep}$. As discussed previously, the parameter $r_D$ only affects the initial conditions whereas the lepton asymmetry, $\xi_{\rm lep}$, affects both the initial conditions, the effective number of neutrino species, as well as $\beta$-decay processes occurring during BBN.  This latter effect alters the primordial light element abundances, so that from measurements of primordial $^4$He and deuterium abundances, we have an independent constraint $|\xi_{\rm lep}| \lesssim 0.03$ at 95\% CL \cite{Dolgov:2002ab,Mangano:2011ip} as discussed in Sec. \ref{sec:decay_scen}. 

In our analysis we try three different priors on $\xi_{\rm lep}$: first we consider the constraints to $\xi_{\rm lep}$ from the CMB only imposing a flat prior on $\xi_{\rm lep}^{2}$ of $0\leq \xi_{\rm lep}^{2}\leq 4$; second we impose the BBN constraint by placing a Gaussian prior on $\xi$ with a mean of zero and a standard deviation of $0.03$; third we consider the case where $\xi_{\rm lep} = 0$, removing the neutrino isocurvature mode. We find that both current CMB measurements by \textit{Planck} and a future cosmic-variance limited experiment (with maximum $\ell = 2200$) are less sensitive to $\xi_{\rm lep}^{2}$ than measurements of the light-element abundances. 

 The observed CMB power spectra can be written in terms of the primordial curvature perturbation power spectrum, $\Delta^2_{\zeta}(k)$, and the photon transfer function $\Theta_{\ell}^{i,X}(k)$ for each initial condition $i$ as 
\begin{align}
C^{XY}_{\ell} = 4\pi\int_0^{\infty} \frac{dk}{k} \Delta_{\zeta}^2(k)& \left[\sum_{i}A_{i\gamma}\Theta_{\ell}^{i,X}(k)\right] \nonumber \\ \times& \left[\sum_{j}A_{j\gamma}\Theta_{\ell}^{j,Y}(k)\right],
\end{align} where $X\in \left\{T,E\right\}$ denotes the relevant observable (CMB temperature or E-mode polarization anisotropy).

The primordial curvature perturbation is given in terms of the amplitude parameter $A_{s}$:
\begin{equation}
\Delta_{\zeta}^{2}(k)\equiv \frac{k^{3}}{2\pi^{2}}P_{\zeta}(k) = A_s \left(\frac{k}{k_0}\right)^{n_s - 1},
\end{equation}
where $P_{\zeta}(k)$ is the dimensional power spectrum of $\zeta$, $A_s$ is the primordial scalar amplitude and $n_s$ is the primordial scalar spectral index and the pivot wavenumber is 
taken to be $k_0 = 0.05\ {\rm Mpc}^{-1}$. 
The amplitude parameters  
\begin{equation}
A_{i\gamma} \equiv \left\{ A_{ad}, A_{c \gamma}, A_{b \gamma}, A_{\nu \gamma}\right\}
\end{equation} are used to set the mixture of adiabatic and isocurvature modes in the CMB Boltzman code \textsc{camb}. It is important to set all these amplitudes correctly in the presence of neutrino isocurvature, as neutrinos contribute to the relativistic energy density at early times, and the neutrino isocurvature density mode is excited in the curvaton model, as we saw in Sec.~\ref{sec:decay_scen}. As discussed in Appendix \ref{appendix}, using the initial perturbation values $\delta_{c}$, $\delta_{b}$, $\delta_{\gamma}$, and $\delta_{\nu}$ for each perturbation mode used in \textsc{camb}, we have that $A_{b\gamma}=S_{b\gamma}/\zeta-R_{\nu}S_{\nu\gamma}/\zeta$, $A_{c\gamma}=S_{c\gamma}/\zeta-R_{\nu}S_{\nu\gamma}/\zeta$, and $A_{\nu\gamma}=3S_{\nu\gamma}R_{\gamma}/4\zeta,$ where $R_{\gamma}\equiv 1-R_{\nu}$ is the fraction of relativistic energy in photons. We apply these relations when running our MCMC chains for each of the curvaton decay-scenarios enumerated in Sec. \ref{sec:decay_scen}, along with Table \ref{tab:curvatoncases} and Eq.~(\ref{eq:neff}).

Before presenting constraints to $r_D$ and $\xi_{\rm lep}$ it is instructive to consider a `model independent' parameterization of the totally correlated (or anti-correlated) isocurvature modes. Fig.~\ref{fig:comp1} gives us a sense of what to expect from this exercise.  First, note that a 20\% contribution from totally correlated CDM isocurvature (blue curves) can produce a deficit of power on large scales while also causing a significant change at around the first peak in the TT power spectrum.  Given the mild tension between the best-fit theoretical power spectrum and the relatively low temperature quadrupole (at the level of a little more than one standard deviation), we expect the data to prefer a slightly negative value for $S_{m \gamma}$. The matter isocurvature also has a significant affect on the TE power spectrum between $100 \lesssim 
\ell \lesssim 500$. 

The introduction of a 10\% contribution from neutrino isocurvature (orange curves) significantly changes the TT power spectrum at nearly all scales as well as 
the TE and EE power spectrum on scales with $\ell \gtrsim 100$.  We therefore expect that the CMB data will be more sensitive to $S_{\nu \gamma}$ than than to $S_{m \gamma}$. 
\begin{figure}[h!]
\begin{center}
\resizebox{!}{8cm}{\includegraphics{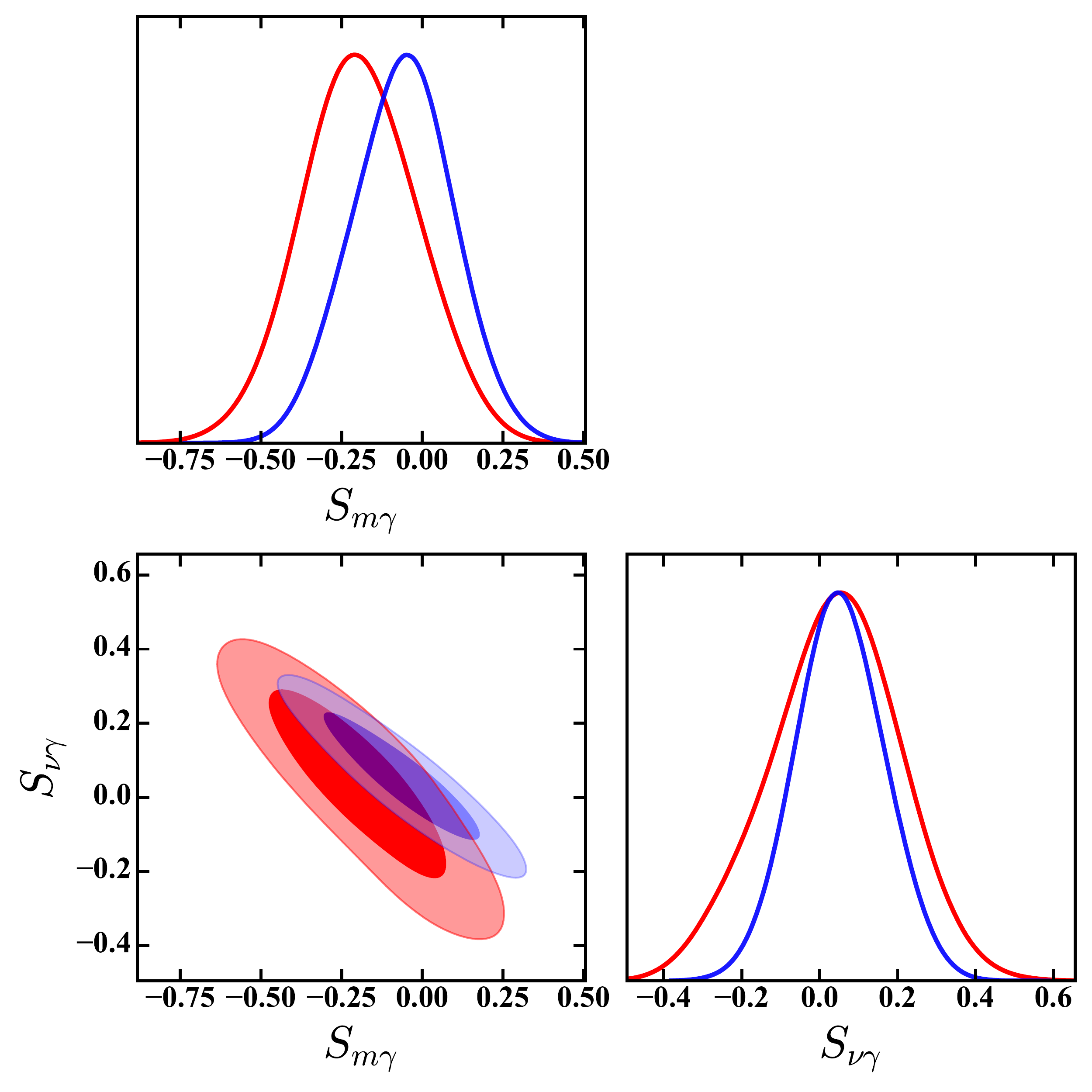}}
\caption{The posteriors for the correlated isocurvature amplitudes $S_{m\gamma}$ and $S_{\nu \gamma}$; the red curves 
show constraints using \textit{Planck} TT+BAO+LowP and the blue curves show constraints using \textit{Planck} TT+BAO+AllP.  Note that at the level of about one standard deviation the LowP case is better fit by a totally anti-correlated matter isocurvature component, which leads to a suppression of power on large angular scales. When all of the polarization data is included this preference is less dramatic.}
\label{fig:corr_iso_1D}
\end{center}
\end{figure} 

The results presented in Fig.~\ref{fig:corr_iso_1D} confirms our expectations: the \textit{Planck} TT+BAO+LowP (red curves) prefers a slightly anti-correlated matter isocurvature amplitude and when all of the polarization data is included (blue curves) the constraints shift towards a purely adiabatic spectrum. We find that the \textit{Planck} TT+BAO+LowP places a constraint $S_{m \gamma } = -0.19\pm 0.18$ and $S_{\nu \gamma} = 0.04^{+0.18}_{-0.16}$ whereas \textit{Planck} TT+BAO+AllP gives $S_{m \gamma } =-0.06\pm 0.16$ and $S_{\nu \gamma} = 0.05\pm 0.11$ at 68\% CL.  

As was noted in Ref.~\cite{Ade:2015xua} the difference between these constraints may be driven by a handful of data points around $\ell \simeq 160$.  This can be seen by eye in Fig~\ref{fig:comp1}: in the top
panel, which shows the TT spectrum, the large-scale residuals 
 are significantly below zero, preferring totally \emph{anti-correlated} matter and 
neutrino density isocurvature (dashed curves); in the TE spectrum there are a few data points around $\ell \simeq 160$ which have residuals significantly above and below zero.  As the  
isocurvature curves show, these data introduce a tension between totally correlated and anticorrelated isocurvature modes.  We note that this tension
may be a significant driver in the difference between the LowP and AllP constraints on isocurvature perturbations, although we do not explore this issue further.

\begin{table}
\begin{tabular}
{lcl}\hline \hline {\rm scenario} & $r_D$ ($A_{m\gamma} = 0$) & $ r_D$ (95\% CL)\\ 
\hline \\
$(b_{\rm by},c_{\rm before},L_{y_{L}})$ & $0.1580^{+0.0042}_{-0.0040}$ & $0.1602^{+0.0051}_{-0.0047}$\\
$(b_{\rm before},c_{\rm by},L_{y_{L}})$ & $0.8373^{+0.0042}_{-0.0043}$ & $0.8492^{+0.0099}_{-0.0096}$ \\
$(b_{\rm by},c_{\rm after},L_{y_{L}})$ & $1$ & $>0.9578$\\
$(b_{\rm after},c_{\rm by},L_{y_{L}})$ & $1$ & $>0.9919$\\
$(b_{\rm by},c_{\rm by},L_{y_{L}})$ & $1$ & $>0.9931$ \\
$(b_{\rm after},c_{\rm after},L_{y_{L}})$ & $1$ & $> 0.9973$\\
\\ \hline \hline 
\end{tabular}
\caption{Constraints to $r_D$ (with $\xi_{\rm lep}^2 = 0$) using \textit{Planck} TT+BAO+LowP to those models which can yield vanishing isocurvature perturbations as seen in any two-point correlation function. Note that the in the scenario $(b_{\rm after},c_{\rm after},L_{y_{L}})$ we quote a constraint to $\chi_D$ which is related to $r_D$ as discussed in more detail in Sec.~\ref{sec:afterafter}.}
\label{compensated_values}
\end{table}
As the data is well-fit by a universe with purely adiabatic perturbations, curvaton scenarios that fit have $r_D$ and $\xi_{\rm lep}$ values that produce adiabatic perturbations.  This immediately eliminates the scenarios $(b_{\rm after}, c_{\rm after}, L_{y_L})$ and $(b_{\rm before}, c_{\rm before}, L_{y_L})$. We also note that the case where $(b_{\rm before},c_{\rm after},L_{y_L})$ will produce a huge isocurvature perturbation, unless $r_D$ exceeds the bounds from non-Gaussianity and $\xi_{\rm lep}$  exceeds the BBN bounds. This scenario is thus ruled out to high significance as well.  We are then left with 18 scenarios which may be consistent with the data. 

Each of the allowed 18 scenarios yield zero isocurvature contributions to CMB power spectra if $A_{m \gamma} = 0$-- i.e., as long as they correspond to a compensated isocurvature mode.  We show the value of $r_D$ in each of these scenarios for which $A_{m\gamma} = 0$ in Table \ref{compensated_values} along with the constraints to $r_D$ when $\xi_{\rm lep} = 0$. 

In addition to running MCMC chains to obtain constraints, we perform a Fisher-matrix analysis to forecast the sensitivity of CMB data to  $\ln{(r_{\rm D})}$ and $\xi_{\rm lep}^{2}$. We include these parameters, as well as the standard $6$ $\Lambda$CDM parameters. We apply the Fisher-matrix formalism as described in Ref. \cite{Eisenstein:1998hr}. In this analysis, we also include a BBN prior on the primordial $\lsup{\rm He}{4}$ abundance, with error $\sigma_{Y_{\rm He}}=0.005$. 

Numerical derivatives are evaluated using a standard two-sided two-point numerical derivative, except for the parameters $\ln{r_{\rm D}}$ and  $\xi_{\rm lep}^{2}$, for which a one-sided seven-pt rule was applied to obtain sufficiently convergent numerical derivatives. Additionally, for $\Omega_{b}h^{2}$, a two-sided seven-pt rule was used to guarantee numerical convergence.  For $\ln{(10^{10}A_{s})}$, the derivative $dC_{\ell}^{XY}/dA_{s}$ was evaluated analytically, as $C_{\ell}^{XY}\propto A_{s}$, obviating the need to compute a numerical derivative for this parameter. These results are used both to verify that our MCMC results for Planck data are reasonable, and to forecast the ideal sensitivity of a cosmic-variance limited CMB polarization experiment to curvaton-generated isocurvature perturbations. 

Fiducial values for $\Omega_{b}h^{2}$, $\Omega_{c}h^{2}$, $\Omega_{\Lambda}$, $A_{s}$, $n_{s}$, and $\tau$ were set to the marginalized means for these parameters in a $\Lambda$CDM-only MCMC run. For the lepton asymmetry, we used $\xi_{\rm lep}^{2}=0$ as the fiducial value. For all curvaton scenarios except ($b_{\rm by},c_{\rm before}, L_{y_{L}}$) and ($b_{\rm before},c_{\rm by}, L_{y_{L}}$), we used the fiducial value
$r_{D}=1$, guaranteeing that the fiducial model has adiabatic perturbations. For the scenarios ($b_{\rm by},c_{\rm before}, L_{y_{L}}$) and ($b_{\rm before},c_{\rm by}, L_{y_{L}}$), we used fiducial values corresponding to zero isocurvature between radiation and non-relativistic matter (i.e., $A_{m\gamma} = 0$), corresponding to the $r_{D}$ values given in the middle column of Table \ref{compensated_values}. 

We now present our constraints to curvaton-decay  scenarios, grouped by the character of their effects on CMB power spectra. We begin by discussing scenarios for which there is non-zero isocurvature unless $\xi=0$ and $r_{D}=\Omega_{i}/\Omega_{m}$, where $i$ denotes baryons or CDM. We then move on to a scenario showing a total degeneracy between $\xi_{\rm lep}^{2}$ and $r_{D}$ at the level of isocurvature amplitudes. We finish by discussing scenarios for which all isocurvature modes vanish when $r_{D}=1$.

\begin{figure}
\resizebox{!}{13.5cm}{\includegraphics{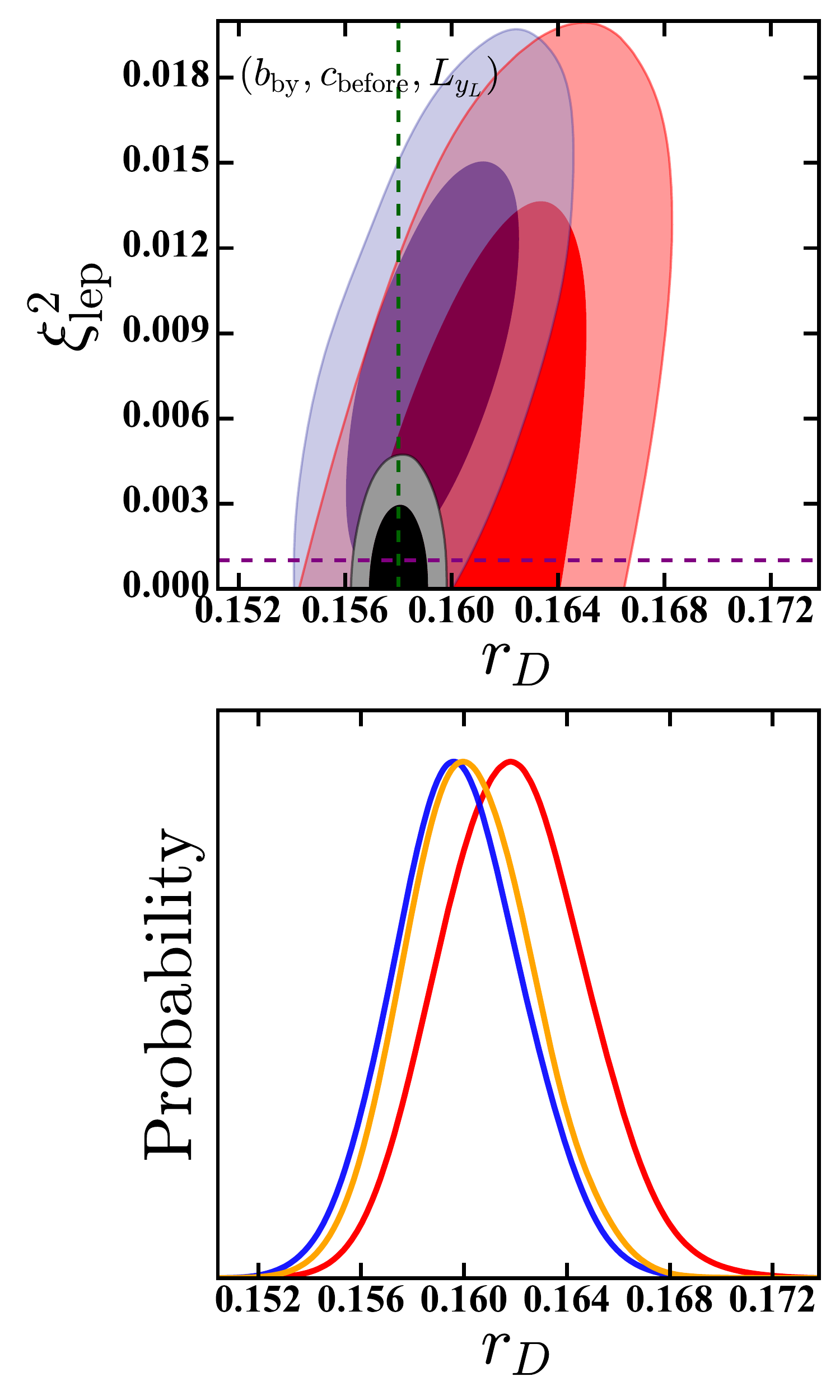}}
\caption{Constraints to $r_{D}$ and $\xi_{\rm lep}^2$ for scenario $(b_{\rm by},c_{\rm before},L_{y_{L}})$.  \emph{Top}: the marginalized 2D constraints to both $r_D$ and $\xi_{\rm lep}$. The red regions show the current constraints using \textit{Planck} TT+BAO+LowP data, the blue regions show constraints using \textit{Planck} TT+BAO+AllP, and the black regions show the projected constraints for a cosmic-variance limited CMB experiment which measures out to 
$\ell_{\rm max} = 2200$, obtained from a Fisher-matrix analysis. In this panel, a flat prior is imposed on $\xi_{\rm lep}^{2}$, as discussed in the text. The dashed vertical green line gives the value of $r_D$ for which the isocurvature is totally compensated (i.e. $A_{m\gamma} = 0$); the dashed horizontal purple line gives the 95\% CL upper limit on $\xi_{\rm lep}^2$ from measurements of the primordial light element abundances. \emph{Bottom}:  marginalized 1D constraints to $r_D$ using \textit{Planck} TT+BAO+LowP under a variety of assumptions for $\xi_{\rm lep}$: flat prior on $\xi_{\rm lep}^2$ (red), BBN-prior on $\xi_{\rm lep}^2$ (blue), and $\xi_{\rm lep}^2 =0$ (orange).}
\label{fig:a6_constraints}
\end{figure} 

\subsection{Constraints to baryon number or CDM production before curvaton decay}
 
The two decay scenarios which produce compensated isocurvature modes are $(b_{\rm by},c_{\rm before},L_{y_{L}})$ and $(b_{\rm before},c_{\rm by},L_{y_{L}})$.  
As shown in Table \ref{tab:curvatoncases}, the isocurvature contribution vanishes (i.e., is purely compensated) when 
$\xi_{\rm lep} = 0$ and $r_D = \Omega_b/\Omega_m$ for $(b_{\rm by},c_{\rm before},L_{y_{L}})$ or $r_D = \Omega_c/\Omega_m$ for $(b_{\rm before},c_{\rm by},L_{y_{L}})$.  In addition to this, if $r_D$ is greater than the previous values the matter isocurvature is anti-correlated with the adiabatic mode, leading to a suppression of the large-scale temperature power spectrum.  As expected, constraints from \textit{Planck} TT+BAO+LowP lead to values of $r_D$ which are slightly \emph{larger} than the purely compensated case, since that leads to a suppression of the large-scale temperature power spectrum.  Marginalizing over $\xi_{\rm lep}^2$ for $(b_{\rm by},c_{\rm before},L_{y_{L}})$ we find that at 95\% CL $r_D = 0.1619^{+0.0055}_{-0.0053}$ and $ \Omega_b/\Omega_m = 0.1580^{+0.0043}_{-0.0041}$; for $(b_{\rm before},c_{\rm by},L_{y_{L}})$ we find that $r_D = 0.856^{+0.015}_{-0.014}$ and $ \Omega_c/\Omega_m =0.8401^{+0.0063}_{-0.0059}$.  

Constraints to $r_D$ in these two scenarios are significantly different when all of the polarization data is included.  In this case marginalizing over $\xi_{\rm lep}^2$ for $(b_{\rm by},c_{\rm before},L_{y_{L}})$ we find that at 95\% CL $r_D = 0.1595^{+0.0044}_{-0.0041}$ and $ \Omega_b/\Omega_m = 0.1570^{+0.0035}_{-0.0033}$; for $(b_{\rm before},c_{\rm by},L_{y_{L}})$ we find that $r_D = 0.853^{+0.015}_{-0.014}$ and $ \Omega_c/\Omega_m =0.8455^{+0.0052}_{-0.0045}$.  We can see that in both scenarios $r_D$ is constrained to be significantly closer to their purely compensated values when all of the polarization data is used. 

The constraint to $\xi_{\rm lep}^2$ in these two scenarios is particularly interesting since the compensated isocurvature leads to a stricter \textit{Planck}/BAO constraint.  Looking at Eq.~(\ref{eq:finalneutrino}) we can see that the smaller $r_D$ the larger the neutrino isocurvature contribution.  This means that the \textit{Planck}/BAO constraints to $\xi_{\rm lep}^2$ for the scenario $(b_{\rm by},c_{\rm before},L_{y_{L}})$ is the most constraining with $\xi_{\rm lep}^2 \leqslant 0.0164$ at 95\% CL as seen in Fig.~\ref{fig:a6_constraints}.  Although this is not competitive with constraints inferred from measurements of the primordial light element abundances \cite{Dolgov:2002ab,Cuoco:2003cu,Guzzo:2003xk,Mangano:2011ip,DiValentino:2014eea}, $\xi_{\rm lep}^2 \leqslant 0.001$ at 95\% CL, it is the tightest constraint to $\xi_{\rm lep}^2$ using only  \textit{Planck}/BAO data.  
\begin{figure}[h!]
\resizebox{!}{13.5cm}{\includegraphics{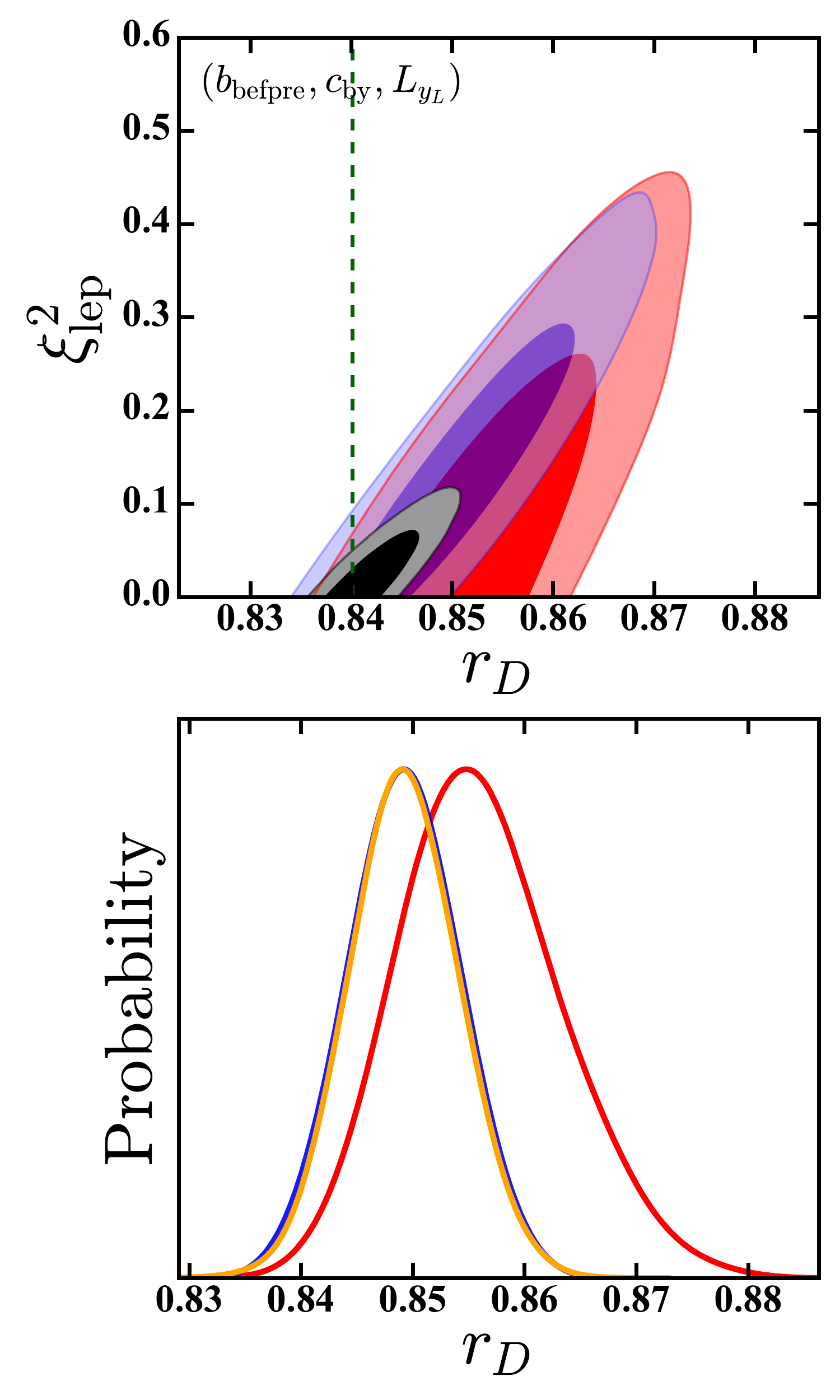}}
\caption{Constraints to $r_{D}$ and $\xi_{\rm lep}^2$ for scenario $(b_{\rm before},c_{\rm by},L_{y_{L}})$.  \emph{Top}: the marginalized 2D constraints to both $r_D$ and $\xi_{\rm lep}$. The red regions show the current constraints using \textit{Planck} TT+BAO+LowP data, the blue regions show constraints using \textit{Planck} TT+BAO+AllP, and the black regions show the projected constraints for a cosmic-variance limited CMB experiment which measures out to 
$\ell_{\rm max} = 2200$, obtained from a Fisher-matrix analysis. In this panel, a flat prior is imposed on $\xi_{\rm lep}^{2}$, as discussed in the text. \emph{Bottom}:  marginalized 1D constraints to $r_D$ using \textit{Planck} TT+BAO+LowP under a variety of assumptions for $\xi_{\rm lep}$: flat prior on $\xi_{\rm lep}^2$ (red), BBN-prior on $\xi_{\rm lep}^2$ (blue), and $\xi_{\rm lep}^2 =0$ (orange).}
\label{fig:a8_constraints}
\end{figure} 

Since the value of $r_D$ is larger in the scenario $(b_{\rm before},c_{\rm by},L_{y_{L}})$ the constraint to $\xi_{\rm lep}^2$ this case is not as restrictive, giving $\xi_{\rm lep}^2 \leqslant 0.368$ at 95\% CL. As shown in Fig.~\ref{fig:Neff}, however, this is more restrictive than the upper limit placed on $\xi_{\rm lep}^2 \leqslant 0.5$ from its contribution to the total radiative energy density of the universe, showing that this constraint is driven by the effect the lepton asymmetry has on neutrino isocurvature perturbations. 
\begin{figure}[h!]
\resizebox{!}{6cm}{\includegraphics{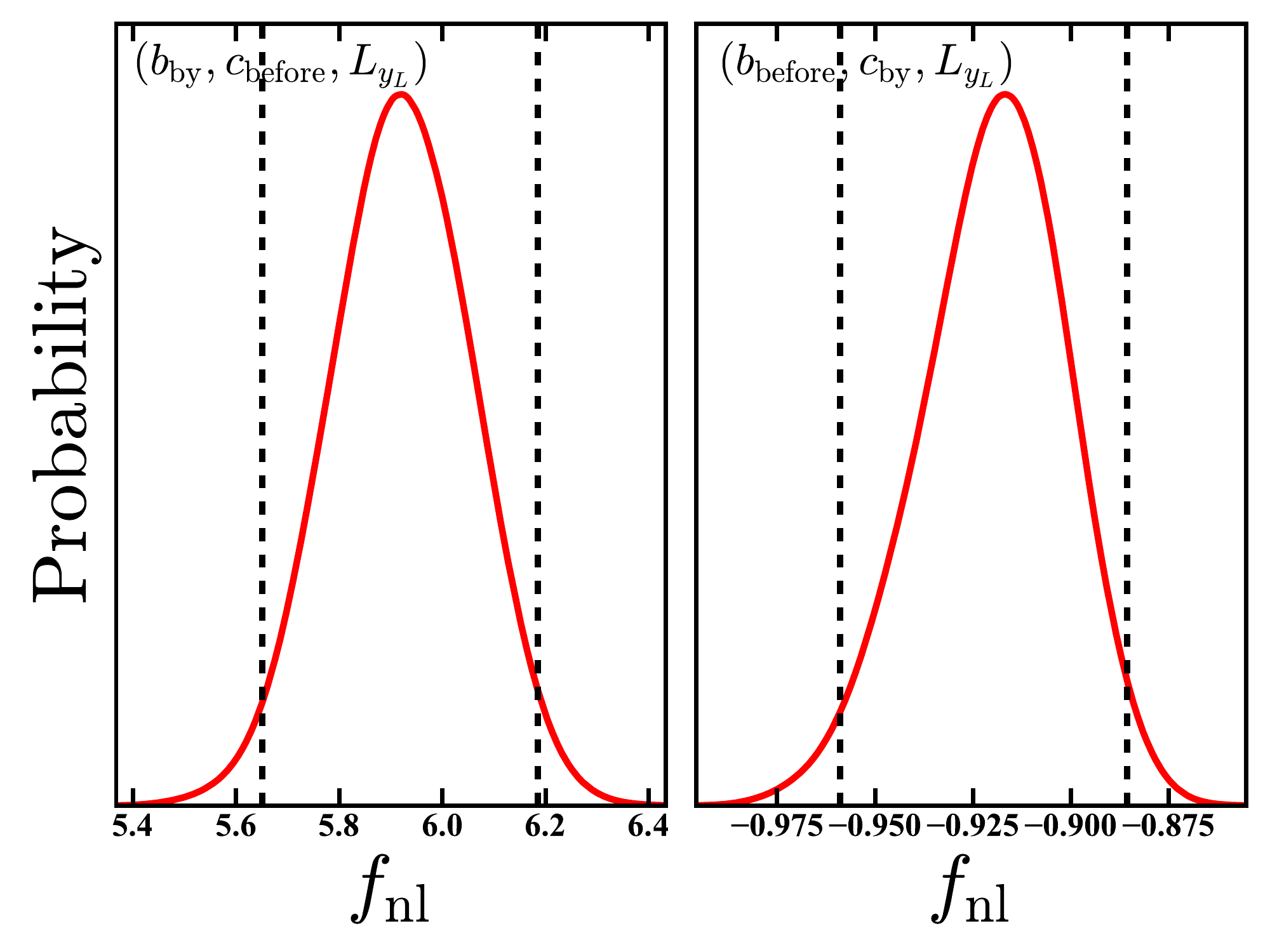}}
\caption{Predicted value of the non-Gaussianity parameter $f_{\rm nl}$ for the scenarios $(b_{\rm by},c_{\rm before},L_{y_{L}})$ and $(b_{\rm before},c_{\rm by},L_{y_{L}})$ for parameter values which are consistent with our limits (on isocurvature and the radiative energy density at decoupling) from \textit{Planck}/BAO data (red).  The vertical dashed lines indicate the 95\% CL range of these predictions.}
\label{fig:fnl_a6_a8}
\end{figure} 

The constraint to $\xi_{\rm lep}^2$ does not change significantly when including all of the polarization data: for scenario $(b_{\rm before},c_{\rm by},L_{y_{L}})$ the constraint becomes $\xi_{\rm lep}^2 \leqslant 0.0165$ and for $(b_{\rm before},c_{\rm by},L_{y_{L}})$ becomes $\xi_{\rm lep}^2 \leqslant 0.348$.  

As shown in the bottom panel of Figs.~\ref{fig:a6_constraints} and \ref{fig:a8_constraints} the marginalized 1D constraint on $r_D$ is fairly insensitive to how we treat $\xi_{\rm lep}^2$. In those panels the red curve shows the constraint arising from flat priors on $\xi_{\rm lep}^{2}$. The blue curve shows the constraint that arises when $\xi_{\rm lep}^2$ has the BBN prior $\xi_{\rm lep}^2 \leq 0.001$ at 95\% CL. The orange curve shows the constraint obtained when we assume $\xi_{\rm lep}^2=0$.  

The values of $r_D$ allowed by \textit{Planck}/BAO data in these scenarios also imply a non-Gaussian signature in the CMB. The predicted level of this signature can be determined through Eq.~(\ref{eq:fnl_predict}).  We show the predicted ranges for the amplitude of this signal, $f_{\rm nl}$, in Fig.~\ref{fig:fnl_a6_a8}. The scenario $(b_{\rm before},c_{\rm by},L_{y_{L}})$ predicts $f_{\rm nl} = 5.92\pm 0.26$ and $(b_{\rm before},c_{\rm by},L_{y_{L}})$ predicts that $f_{\rm nl} = -0.919^{+0.034}_{-0.040}$ at 95\% CL. 
Current data impose the constraint $f_{\rm nl} = 2.5 \pm 5.7$ \cite{Ade:2015ava}.  The scenario $(b_{\rm by},c_{\rm before},L_{y_{L}})$ implies a particularly large $f_{\rm nl}$ value, which could be sensitively tested using measurements of scale-dependent bias in future galaxy surveys \cite{Dalal:2007cu,dePutter:2014lna} or measurements of the matter bispectrum from high-redshift $21$-cm experiments \cite{Cooray:2004kt,Pillepich:2006fj,Munoz:2015eqa}. The scenario $(b_{\rm before},c_{\rm by},L_{y_{L}})$, which makes more modest predictions, could be tested with high-redshift $21$-cm experiments \cite{Cooray:2004kt,Pillepich:2006fj,Munoz:2015eqa}.

Future CMB measurements will greatly improve upon these constraints.  As shown by the black ellipses in Figs.~\ref{fig:a6_constraints} and \ref{fig:a8_constraints} a cosmic-variance limited CMB experiment which measures both the temperature and polarization power-spectrum out to $\ell_{\rm max} = 2200$ will give a factor of 4.3 increase in sensitivity to $\xi_{\rm lep}^2$ and a factor of 3.5 increase in sensitivity to $r_D$ for the scenario $(b_{\rm by},c_{\rm before},L_{y_{L}})$ and a factor of 11 increase in sensitivity to $\xi_{\rm lep}^2$ and a factor of 4 increase in sensitivity to $r_D$ for the scenario $(b_{\rm before},c_{\rm by},L_{y_{L}})$.  Note that even with the increased sensitivity,  CMB/BAO measurements of $\xi_{\rm lep}^2$ are still not as sensitive as measurements of the primordial light-element abundances. 

\subsection{Constraints to baryon and CDM production after curvaton decay}
\label{sec:afterafter}
\begin{center}
\begin{figure*}
\resizebox{!}{5cm}{\includegraphics{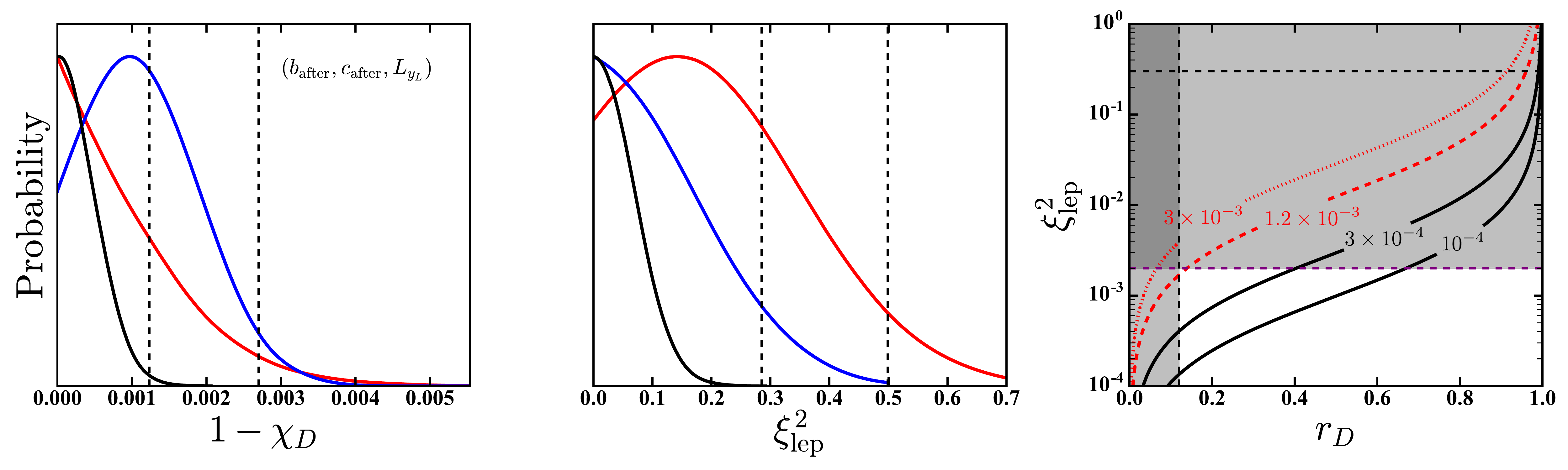}}
\caption{Marginalized 1D constraints to a scenario in which lepton number is produced by curvaton decay, while baryon number and CDM are produced after curvaton decay.  For the left and middle panels the red curve shows constraints using \textit{Planck} TT+BAO+LowP, the blue curve shows constraints using \textit{Planck} TT+BAO+AllP,
and the black curve shows projected constraints for a cosmic-variance limited CMB experiment which measures out to 
$\ell_{\rm max} = 2200$, obtained from a Fisher-matrix analysis. The vertical dashed lines indicate the 95\% CL upper limit to each parameter using the \textit{Planck} TT+BAO+LowP data.  \emph{Left hand panel}: the 1D marginalized posterior 
for $1-\chi_D$, where $\chi_D$ is defined in Eq.~(\ref{eq:defchiD}).  \emph{Middle panel} the 1D marginalized posterior on $\xi_{\rm lep}^2$ from CMB/BAO observations only.  
\emph{Right-hand panel}: a contour plot showing the relationship between $\chi_D$, $r_D$, and $\xi_{\rm lep}^2$.  The dotted red contour shows the 68\% CL upper limit on $1-\chi_D$ from \textit{Planck} TT+BAO+LowP; the dashed red contour shows the 95\% CL upper limit on $1-\chi_D$ from \textit{Planck} TT+BAO+LowP.  The vertical dashed line shows the 95\% CL lower limit on $r_D$ from constraints to the level of non-Gaussianity in the CMB; the horizontal dashed lines show the 95\% CL upper limits on $\xi_{\rm lep}^2$ from the \textit{Planck} TT+BAO+LowP data (black) and measurements of the primordial light element abundances (red).  The shaded region is currently ruled out at 95\% CL.}
\label{fig:a1_constraints}
\end{figure*} 
\end{center}

In the scenario where both the baryon number and CDM are produced after curvaton decay, while lepton number is produced by its decay, the initial conditions are completely determined by the level of neutrino isocurvature alone, as shown in Table \ref{tab:curvatoncases}.  Looking at Eq.~(\ref{eq:finalneutrino}) we can see this produces a perfect degeneracy between $r_D$ and $\xi_{\rm lep}^2$: the level of isocurvature can be made to be arbitrarily small for any value of $r_D\leqslant1$ with a small-enough value for $\xi_{\rm lep}^2$.  In order to determine the allowed region of parameter space, it is convenient to define a new parameter, $\chi_D$:
\begin{equation}
\frac{1}{\chi_D}-1 \equiv \frac{\xi_{\rm lep}^2}{\pi^2} \left(\frac{1}{r_D} -1\right).
\label{eq:defchiD}
\end{equation}
The constraints to $\chi_D$ and $\xi_{\rm lep}^2$ are shown in Fig.~\ref{fig:a1_constraints}. 

As discussed in Sec.~\ref{sec:lep}, even in the absence of neutrino isocurvature, \textit{Planck}/BAO data impose the constraint $\xi_{\rm lep}^{2}\leqslant 0.49$ (at 95\% CL), due to the effect of $\xi_{\rm lep}^{2}$ on $N_{\rm eff}$. We have also seen that measurements of the primordial light element abundances further constrain $\xi_{\rm lep}^2 \leqslant 0.001$ at 95\% CL.  The unshaded region of the rightmost panel of Fig.~\ref{fig:a1_constraints} shows the currently allowed region of the $r_D$-$\xi_{\rm lep}^2$ parameter space in this scenario. 

The \textit{Planck} TT+BAO+LowP data places the constraint $1-\chi_D \leqslant 0.0027$ and $\xi_{\rm lep}^2 \leqslant 0.5$ at 95\% CL.  These data have a slight preference for non-zero $\xi_{\rm lep}^2$ due to its additional contribution to the radiative energy density.  This preference has been seen in previous analyses \cite{Smith:2011es,Hou:2011ec,Ade:2015xua}.  When all the polarization data are used, the preference for a non-zero $\xi_{\rm lep}^2$ disappears but is replaced by a slight preference for $\chi_D<1$, as can be seen in the blue curves in the left panel of Fig.~\ref{fig:a1_constraints}. In this case we have $1-\chi_D \leqslant 0.0025$ and $\xi_{\rm lep}^2 \leqslant 0.33$ at 95\% CL.

Given that any value of $r_D$ is consistent with the \textit{Planck}/BAO data, this scenario does not make a specific prediction for a level of non-Gaussianity.  Instead, current ($f_{\rm nl} = 2.5 \pm 5.7$ \cite{Ade:2015ava}) allow us to conclude that $r_D \geqslant 0.12$ at 95\% CL.  This constraint is shown in the left-hand panel of Fig.~\ref{fig:a1_constraints} as the vertical dashed line.  

Future measurements of the CMB will more sensitive to this curvaton-decay scenario, as shown by the black curves in the left and center panel of Fig.~\ref{fig:a1_constraints}. Using a Fisher-matrix analysis, we find that a cosmic-variance limited experiment which measures both the temperature and polarization power-spectrum out to $\ell = 2200$ will be 4 times more sensitive to $\chi_D$ and 3 times more sensitive to $\xi_{\rm lep}^2$. 

\subsection{Constraints to remaining scenarios}
As shown in Table \ref{tab:curvatoncases}, unlike the other cases considered, these three scenarios yield purely adiabatic initial conditions when $r_D = 1$.  
This has important implications for \textit{Planck}/BAO constraints to $\xi_{\rm lep}^2$ in these scenarios.  From Eq.~(\ref{eq:finalneutrino}), it is clear that the level of neutrino isocurvature in these models is negligible. As a result, the sensitivity of \textit{Planck}/BAO data to 
 $\xi_{\rm lep}^2$ comes solely from its contribution to the total radiative energy density of the universe. 
 
\begin{center}
\begin{figure*}
\resizebox{!}{5cm}{\includegraphics{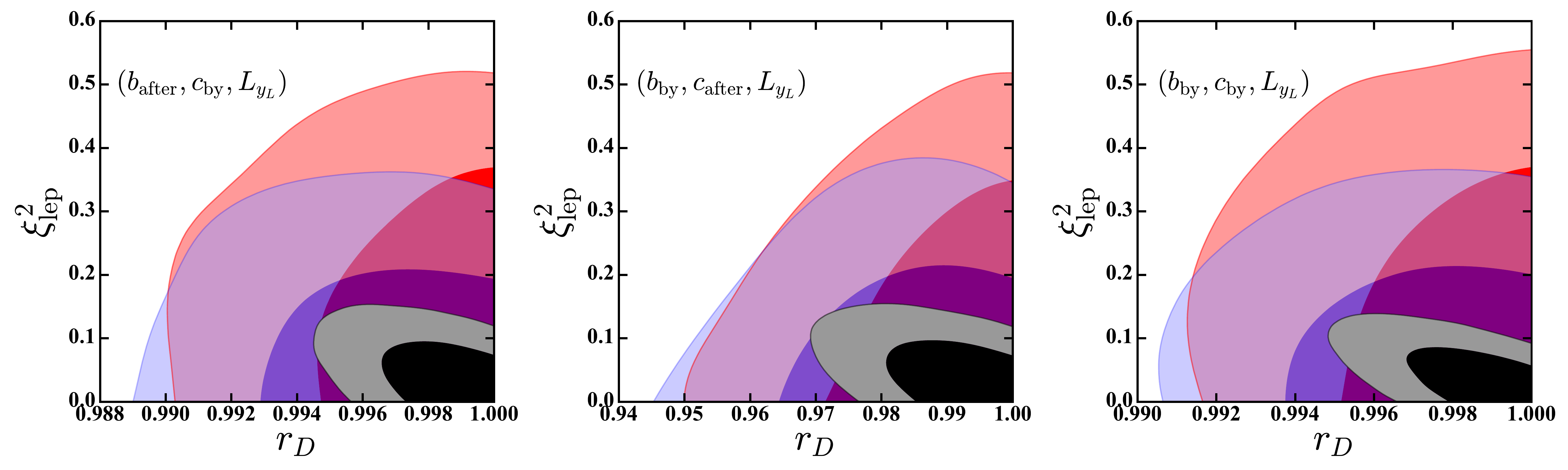}}
\caption{Constraints to the three cases where the baryon number and/or CDM is produced after the curvaton decays.  The red regions show the current constraints using \textit{Planck} TT+BAO+LowP data, the blue regions show constraints using \textit{Planck} TT+BAO+AllP, and the black regions show the projected constraints for a cosmic-variance limited CMB experiment which measures out to 
$\ell_{\rm max} = 2200$, obtained from a Fisher-matrix analysis.  In all cases the inner contour corresponds to 68\% CL and the outer contour corresponds to 95\% CL.  In this panel, a flat prior is imposed on $\xi_{\rm lep}^{2}$, as discussed in the text. In these three cases the initial conditions are purely adiabatic when $r_D=1$.}
\label{a245_constraints}
\end{figure*} 
\end{center}

 This expectation is borne out in Fig.~\ref{a245_constraints} since in all $3$ scenarios give nearly the same 95\% CL upper limit from the \textit{Planck}/BAO data for $\xi_{\rm lep}^2$: for $(b_{\rm after},c_{\rm by}, L_{y_L})$ $\xi_{\rm lep}^2 \leqslant 0.42$; for $(b_{\rm by},c_{\rm after}, L_{y_L})$ $\xi_{\rm lep}^2 \leqslant 0.40$; for $(b_{\rm by},c_{\rm by}, L_{y_L})$ $\xi_{\rm lep}^2 \leqslant 0.44$. When using all of the polarization data (blue regions in Fig.~\ref{a245_constraints}) the sensitivity to $\xi_{\rm lep}^2$ is significantly improved. In all $3$ cases, the 95\%-confidence upper limit to $\xi_{\rm lep}^2$ is a factor of $\simeq 0.75$ of its value for less complete polarization data.
The constraint to $r_D$ in each scenario varies because of the specific pre-factor generated in each case.  Looking at Table \ref{tab:curvatoncases} we can see that the overall matter isocurvature in scenario $(b_{\rm by},c_{\rm after}, L_{y_L})$ is suppressed by the small factor $\Omega_b/\Omega_m \simeq 0.15$.  Because of this we expect the constraint to $r_D$ in that case to be the least restrictive.  The factor $\Omega_c/\Omega_m \simeq 0.8$ appears in the expression for the isocurvature amplitude in the scenario $(b_{\rm after},c_{\rm by}, L_{y_L})$, leading to a moderate suppression of the matter isocurvature.  Finally, since the scenario $(b_{\rm by},c_{\rm by}, L_{y_L})$ contains no suppression we expect the most restrictive constraint on $r_D$ to occur in this case.  All of these expectations are borne out, as shown in Fig.~\ref{a245_constraints}. The 95\% CL lower limits for $(b_{\rm after},c_{\rm by}, L_{y_L})$ is $r_D \geqslant 0.992$. For $(b_{\rm by},c_{\rm after}, L_{y_L})$, the limit is $r_D \geqslant 0.963$. Finally for $(b_{\rm by},c_{\rm by}, L_{y_L})$, the limit is $r_D \geqslant 0.993$. When using all of the polarization data (blue regions in Fig.~\ref{a245_constraints}) the sensitivity to $r_D$ is nearly unchanged. 

\begin{figure}
\resizebox{!}{6cm}{\includegraphics{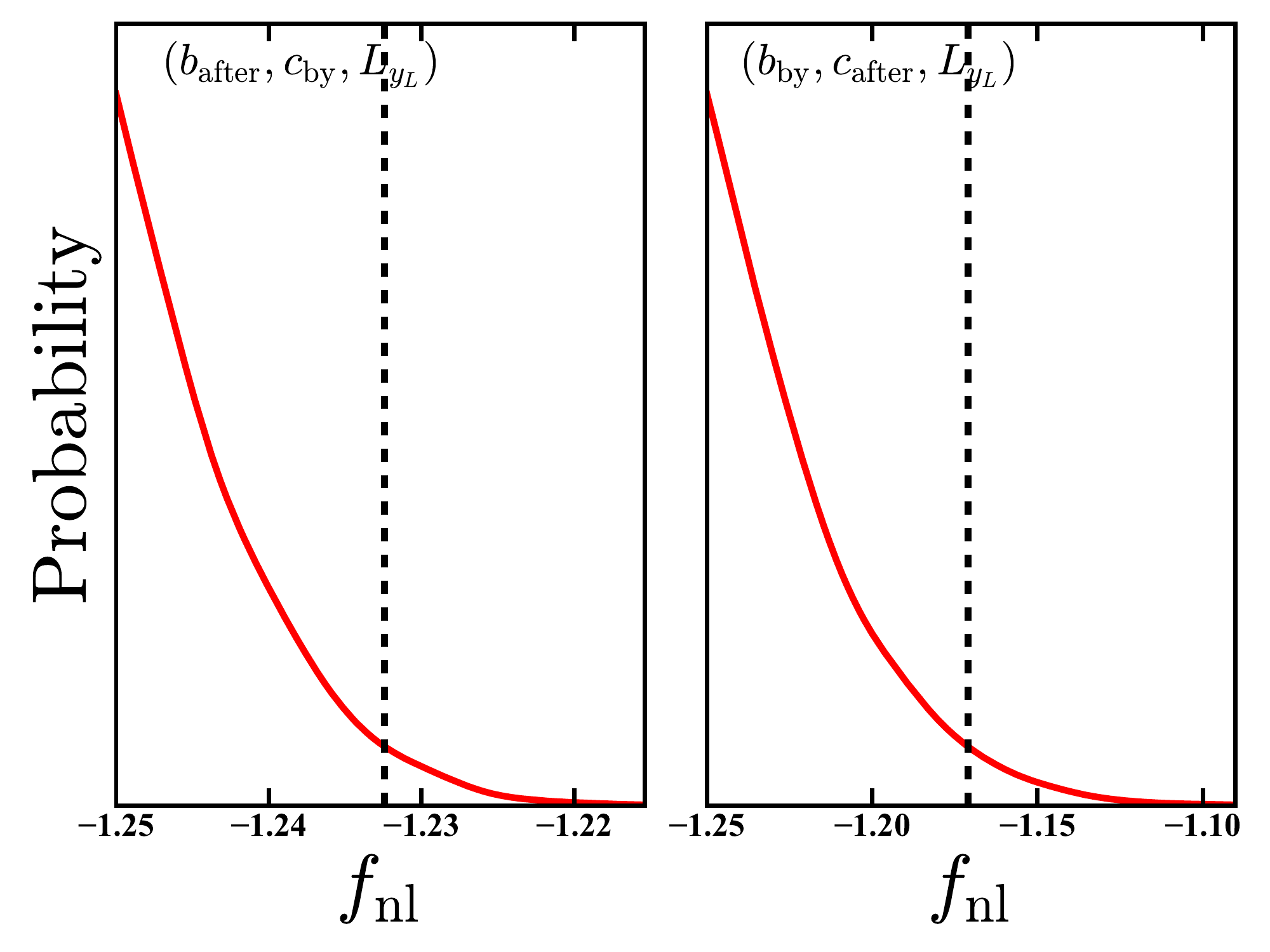}}
\caption{Predicted value of the non-Gaussianity parameter $f_{\rm nl}$ for the scenarios $(b_{\rm after},c_{\rm by},L_{y_{L}})$ and $(b_{\rm by},c_{\rm after},L_{y_{L}})$ for parameter values which are consistent with our limits (on isocurvature and the radiative energy density at decoupling) from \textit{Planck}/BAO data (red).  The vertical dashed lines indicate the 95\% CL range of these predictions. The results for the $(b_{\rm by},c_{\rm by},L_{y_{L}})$ scenario are indistinguishable from those for the $(b_{\rm after},c_{\rm by},L_{y_{L}})$ scenario.}
\label{fig:fnl_a245}
\end{figure}

The values of $r_D$ in these scenarios which are consistent with the \textit{Planck}/BAO lead to a non-Gaussian signal.  The predicted level of this signal can be determined through Eq.~(\ref{eq:fnl_predict}).  Note that the predicted values of $f_{\rm nl}$ are bounded from below, since when $r_D = 1$ we have $f_{\rm nl} = -1.25$.  We show the predicted ranges for the amplitude of this signal, $f_{\rm nl}$, in Fig.~\ref{fig:fnl_a245}.  Since the upper limit on $r_D$ for the scenarios $(b_{\rm after},c_{\rm by}, L_{y_L})$ and $(b_{\rm by},c_{\rm by}, L_{y_L})$ is more restrictive, the 95\% CL upper limit on the \emph{predicted} level of non-Gaussianity in these scenarios is more restrictive with $-1.25 \leqslant f_{\rm nl}\leqslant -1.23$ whereas for $(b_{\rm by},c_{\rm after}, L_{y_L})$, we have $-1.25 \leqslant f_{\rm nl}\leqslant -1.17$. Current Planck data indicate that $f_{\rm nl} = 2.5 \pm 5.7$ \cite{Ade:2015ava}, and so both of these scenarios are consistent with current constraints to primordial non-Gaussianity.
These $f_{\rm nl}$ values could, however, be tested using future measurements of the matter bispectrum from high-redshift $21$-cm experiments \cite{Cooray:2004kt,Pillepich:2006fj,Munoz:2015eqa}.

Future CMB measurements will greatly improve these constraints.  As shown by the black ellipses in Fig.~\ref{a245_constraints} a cosmic-variance limited CMB experiment which measures both the temperature and polarization power-spectrum out to $\ell_{\rm max} = 2200$ will give a factor of 3.5 increase in sensitivity to $\xi_{\rm lep}^2$ and a factor of 2 increase in sensitivity to $r_D$ for each of the three scenarios considered in this subsection.  

\section{Conclusions}
\label{sec:conclusions}

The curvaton scenario presents a rich and interesting alternative to standard single-field slow-roll inflationary models of early universe physics.  There are 
27 curvaton-decay scenarios, distinguished by whether baryon number, lepton number, and CDM are produced before, by, or after curvaton decay. Of these, 18 are currently allowed by CMB and large-scale structure measurements.  

Sensitivity to $r_D$, a parameter describing the curvaton energy-density, comes from the effects of non-adiabatic initial conditions on the CMB as well as the introduction of non-Gaussian statistics. Constraints on $\xi_{\rm lep}^2$, the lepton-number chemical potential, come from the effects of non-adiabatic initial conditions on the CMB as well as its contribution to the total radiative energy density.  

We compared predictions for CMB anisotropy power spectra in these $18$ scenarios with \textit{Planck} CMB measurements and the location of the BAO peak.  The CMB data is divided between large-scale and small-scale measurements.  As noted in Refs.~\cite{Ade:2015xua,Ade:2015lrj} the inclusion of the small-scale polarization data significantly improves sensitivity to isocurvature perturbation.  We find that, when the small-scale polarization data is also used to measure the curvaton scenario parameters, the improved sensitivity is less significant, due, in part, to degeneracies between parameters.

For cases where $r_{D}=1$ restores totally adiabatic perturbations, we find limits of $r_{D}>0.96-0.997$ at 95\% CL, depending on the precise decay-scenario. In these cases, constraints to $\xi_{\rm lep}^2$ are primarily driven by its effect on the relativistic energy-density with $\xi_{\rm lep}^2\leqslant 0.5$ at 95\% CL. These scenarios (with the exception of the decay scenario in which both CDM and baryons are produced after curvaton decay) predict $f_{\rm nl}\sim -1.25$, a value which could be tested by future high-redshift $21$-cm surveys \cite{Cooray:2004kt,Pillepich:2006fj,Munoz:2015eqa}. When both CDM and baryons are produced after curvaton decay, $r_D$ and $\xi_{\rm lep}^2$ are completely degenerate and no specific prediction for $f_{\rm nl}$ can be made. 

The most interesting cases from an observational point of view are those in which baryon number is produced by curvaton decay, and CDM before, or vice-versa. The data then require that $r_{D}=0.160 \pm 0.004$ or $r_{\rm D}=0.850\pm 0.009$ at $95\%$-confidence for these two cases, respectively. This window results from the requirement that the baryon and CDM isocurvature fluctuations nearly cancel, making testable predictions for future experiments. 

First of all, there must be a compensated isocurvature perturbation between baryons and CDM to obtain a small overall isocurvature amplitude \cite{Gordon:2009wx}. In the curvaton model, this CIP must be totally correlated with $\zeta$, and a future CMB experiment (such as CMB Stage-IV \cite{Abazajian:2013oma}) could test the scenario in which baryon number is generated by curvaton decay and CDM before \cite{He:2015msa}. The scenario in which CDM is generated by curvaton decay and baryon number before is inaccessible to CMB searches for compensated isocurvature perturbations \cite{He:2015msa}.

Second of all, in these decay scenarios the perturbations are non-Gaussian. The non-Gaussian signal is larger than in the cases where $r_D = 1$, since the limit of adiabatic perturbations corresponds to $r_{D}<1$ in these scenarios. We find that when baryon number is produced by curvaton decay and CDM before, the parameter values allowed by the CMB power spectra/BAO data predict that $f_{\rm nl} = 5.92 \pm 0.26$. This is still within the current limits to $f_{\rm nl}$ from the CMB bispectrum and may be detected by future galaxy surveys \cite{Dalal:2007cu} (through scale-dependent bias) and high-redshift $21$-cm experiments \cite{Cooray:2004kt,Pillepich:2006fj,Munoz:2015eqa}. If CDM is produced by and baryon number is produced before curvaton decay the model predicts $f_{\rm nl} = -0.919^{+0.034}_{-0.04}$, detection is more challenging, but perhaps possible with 
high-redshift $21$-cm experiments \cite{Cooray:2004kt,Pillepich:2006fj,Munoz:2015eqa}.

If lepton number is produced by curvaton decay, the requirement that neutrino isocurvature perturbations satisfy constraints imposes a limit on  $\xi_{\rm lep}$. If baryon number is produced by curvaton decay, CDM after, and lepton number by the decay, the \textit{Planck} data require $\xi_{\rm lep}\leq 0.13$, much tighter than the constraint to $\xi_{\rm lep}$ obtained from the overall radiation energy density at the surface of last scattering.

Conservatively speaking, future CMB experiments may bring an additional factor of $\sim 3$ improvement in sensitivity to deviations of $r_{D}$ from values consistent with purely adiabatic fluctuations. Depending the precise character of small-scale polarized foregrounds \cite{Calabrese:2014gwa}, primary CMB polarization anisotropies could be measured at multipole scales as high as $\ell\sim 4000$, further improving sensitivity to curvaton-generated isocurvature. Furthermore, primordial initial conditions should have an imprint on the shape of the BAO peak, going beyond the simple location of the peak in real space. This effect could yield an additional test of the curvaton model, if it can be disentangled from redshift-space distortions and nonlinearities.
\begin{acknowledgments}
We acknowledge useful conversations with W.~Hu, C.~He, and R.~Hlo\v{z}ek.~We thank Y.~Ali-Ha\"{i}moud for useful conversations and a thorough reading of the manuscript.~DG is funded at the University of Chicago by a National Science Foundation Astronomy and Astrophysics Postdoctoral Fellowship under Award NO. AST-1302856.~We acknowledge support from the Kavli Foundation and its founder Fred Kavli and by U.S. Dept. of Energy contract DE-FG02-13ER41958, which made possible visits by TLS to the Kavli Institute for Cosmological Physics at the University of Chicago. \end{acknowledgments}
\begin{appendix}
\section{Derivation of relation between isocurvature amplitude and initial mode amplitude \label{appendix}}
We now derive relationships between the mode amplitudes $A_{i\gamma}$ used in \textsc{camb} and the physical isocurvature amplitudes $S_{i\gamma}$ predicted by the curvaton-decay scenarios in Table \ref{tab:curvatoncases}. In terms of the curvature perturbation on hypersurfaces of constant single-species energy density ($\zeta_{i}$), we have 
\begin{equation}
S_{i\gamma} \equiv 3\left(\zeta_i - \zeta_\gamma\right) = - 3 \mathcal{H} \left(\frac{\delta \rho_i}{\rho'_i} - \frac{ \delta \rho_\gamma}{\rho'_{\gamma}} \right),
\end{equation} where $\delta \rho_i = \rho_i \Delta_i$, the prime indicates a derivative with respect to conformal time and $\mathcal{H}$ is the conformal Hubble rate (and $\mathcal{H} = a H$). 

The continuity equation dictates that 
 \begin{equation}
 \dot{\rho}_i = - 3 H \rho_i(1+w_i) \rightarrow \rho'_i = - 3\mathcal{H} \rho_i (1+w_i).
 \end{equation}
 Therefore we can write the isocurvature perturbation in terms of the relative energy density perturbation $\Delta_i$:
\begin{equation}
S_{i\gamma} = \frac{1}{1+w_i}\Delta_i - \frac{3}{4} \Delta_\gamma.\label{eq:eos_ent}
\end{equation}
We can now see that adiabatic initial conditions take the usual form
\begin{equation}
\Delta_c = \Delta _b = \frac{3}{4} \Delta_\gamma = \frac{3}{4} \Delta_\nu.
\end{equation}

Now we can also see how to translate the conditions given here to the initial conditions specified in a Boltzmann solver such as CAMB. For example, with CDM isocurvature we have
\begin{widetext}
\begin{equation}
S_{c\gamma} = \Delta_c - \frac{3}{4} \Delta_\gamma = A_{c\gamma}\left[1- 2 \Omega_{c,0} \tau + 3 \Omega_{c,0} \tau^2 - \frac{3}{4} \left(-\frac{8}{3} \Omega_{c,0}\tau + 4 \Omega_{c,0} \tau^2\right)\right] = A_{c\gamma} \zeta,
\end{equation}
\end{widetext}where we have applied the super-horizon power series solution for the CDM isocurvature mode from Ref. \cite{Bucher:2000kb} and then evaluated it at initial conformal time $\tau=0$. This means that if this mode is excited with an amplitude $A_{c\gamma}$ (relative to the adiabatic mode) that $S_{c\gamma} = A_{c\gamma} \zeta$. 

When we excite multiple isocurvature modes, then the overall isocurvature is the linear combination of each mode.  Exciting both the CDM (with amplitude $A_{c\gamma}$) and baryon isocurvature (with amplitude $A_{b\gamma}$) modes leads to 
\begin{eqnarray} 
S_{c\gamma} &=& \Delta_c - \frac{3}{4} \Delta_\gamma = A_{c\gamma} \zeta, \\
S_{b\gamma} &=& \Delta_b - \frac{3}{4} \Delta_\gamma = A_{b\gamma} \zeta.
\end{eqnarray}
Things get more interesting when we consider the excitation of both matter and neutrino density isocurvature.  
The linear combination of CDM, baryon, and neutrino density isocurvature gives initial density contrasts (applying the power-series solutions from Ref. \cite{Bucher:2000kb} again):
\begin{eqnarray}
\Delta_\gamma &=& A_{c\gamma} \Delta_{\gamma,0} + \frac{R_b}{R_c} A_{b\gamma} \Delta_{\gamma,0} - A_{\nu\gamma} \frac{R_{\nu}}{R_\gamma}, \\
\Delta_c &=& A_{c\gamma} \left(1+ \frac{3}{4} \Delta_{\gamma,0}\right) + \frac{3}{4}\frac{R_b}{R_c}\Delta_{\gamma,0} A_{b\gamma},\\
\Delta_b &=& \frac{3}{4} \Delta_{\gamma,0} A_{c\gamma} + \left(1+\frac{3}{4} \Delta_{\gamma,0} \frac{R_b}{R_c}\right) A_{b\gamma}, \\
\Delta_\nu &=& A_{c\gamma} \Delta_{\gamma,0} + \frac{R_b}{R_c} \Delta_{\gamma,0} A_{b\gamma} + A_{\nu\gamma},
\end{eqnarray}
where $\Delta_{\gamma,0}$ is a constant, $R_c = \rho_c/(\rho_c + \rho_b)$, $R_b = \rho_b/(\rho_c + \rho_b)$, and $R_{\nu} = \rho_{\nu}/(\rho_\gamma + \rho_\nu)$.

Applying Eq.~(\ref{eq:eos_ent}), we then find that
\begin{eqnarray}
S_{c\gamma}/\zeta &=& \left(A_{c\gamma} + \frac{3}{4} \frac{R_\nu}{R_\gamma} A_{\nu\gamma}\right),\\
S_{b \gamma}/\zeta &=& \left( A_{b\gamma} + \frac{3}{4} \frac{R_\nu}{R_\gamma} A_{\nu\gamma}\right),\\
S_{\nu \gamma}/\zeta &=& A_{b\gamma} R_b + A_{c\gamma} R_c + \frac{3}{4} \frac{R_\nu}{R_\gamma}.
\end{eqnarray}

Solving this set of equations for the initial condition amplitudes in terms of the isocurvature amplitudes, we obtain
\begin{eqnarray}
A_{c\gamma} &=& S_{c\gamma}/\zeta + (R_\gamma -1) S_{\nu\gamma}/\zeta, \\
A_{b\gamma} &=& S_{b\gamma}/\zeta +(R_\gamma -1) S_{\nu\gamma}/\zeta, \\
A_{\nu\gamma} &=& \frac{3}{4} R_\gamma S_{\nu\gamma}/\zeta.\label{eq:avals}
\end{eqnarray}

\end{appendix}

\bibliography{chen_spires}

\end{document}